\documentclass[iop]{emulateapj}
\usepackage{color}
\usepackage{graphicx}
\usepackage{longtable}
\usepackage{lineno}
\usepackage{draftcopy}
\usepackage{threeparttable}
\usepackage{multirow}
\linenumbers

\slugcomment{version \today: fm}
\shorttitle{}
\shortauthors{}

\begin{document}
\title{The Two-Dimensional Projected Spatial Distribution of Globular Clusters: I. Method and 
application to NGC4261.}
\author{R.~D'Abrusco\altaffilmark{1}, G.~Fabbiano\altaffilmark{1}, J.~Strader\altaffilmark{2}, 
A.~Zezas\altaffilmark{1,3,4}, S.~Mineo\altaffilmark{1}, T.~Fragos\altaffilmark{1}, P.~Bonfini\altaffilmark{3}, 
B.~Luo\altaffilmark{5}, D.-W.~Kim\altaffilmark{1}, A.~King\altaffilmark{6}}

\altaffiltext{1}{Harvard-Smithsonian Astrophysical Observatory, 60 Garden Street, Cambridge, MA 02138, USA}
\altaffiltext{2}{Department of Astronomy, Michigan State University, 567 Wilson Road, East Lansing, 
MI  48824-2320, USA}
\altaffiltext{3}{Physics Department and Institute of Theoretical and Computational Physics, 
University of Crete, 71003 Heraklion, Crete, Greece}
\altaffiltext{4}{Foundation of Research and Technology, 71003, Heraklion, Crete, Greece}
\altaffiltext{5}{Department of Astronomy \& Astrophysics, 525 Davey Lab, The Pennsylvania State University, 
University Park, PA 16802, USA}
\altaffiltext{6}{Department of Physics \& Astronomy, University of Leicester, Leicester, UK}

\begin{abstract}

We present a new method for the determination of the two-dimensional (2D) projected spatial distribution 
of globular clusters (GCs) in external galaxies. This method is based on the K-Nearest Neighbor 
density estimator of~\cite{dressler1980}, 
complemented by Monte-Carlo simulations to establish the statistical significance of the results. We apply this 
method to NGC4261, a ``test galaxy'' where significant 2D anisotropy in the GC distribution has been 
reported. We confirm that the 2D distribution of GC is not azimuthally isotropic. 
Moreover, we demonstrate that the 2D distribution departures from the average GC radial distribution 
results in highly significant spiral-like or broken shell features. Overall, the same perturbations are 
found in ``red'' and ``blue'' GCs, but with some differences. In particular, we observe a central feature, 
roughly aligned with the minor axis of NGC4261, composed of red and most luminous GCs. Blue and 
fainter GCs are more frequent at large radial distances and follow the spiral-like features of the overall
density structure. These results suggest a complex merging history for NGC4261.

\end{abstract}


\section{Introduction}
\label{sec:intro}

A considerable body of work has been accumulated in the past two decades on the properties of 
GCs populations in elliptical galaxies (see review by~\citealt{brodie2006}). 
The studies available in the literature that use both Hubble Space Telescope (HST) and larger 
ground based telescopes, have established the 
existence of red (metal rich) and blue (metal poor) subpopulations of GCs in most galaxies. The radial 
distributions of these populations differ, with the red/metal rich GCs being more centrally concentrated
than blue GCs. 
These results suggest different formation scenarios for the two GC subpopulations, which may 
reflect the formation history of their parent galaxy. \cite{brodie2006}, in particular, argue that 
blue GCs may have formed in low-mass dark matter halos in the early universe, while the red GCs 
were built in subsequent dissipative buildup of the parent galaxy. Other studies have 
investigated possible formation mechanisms for the families of low-metallicity and high-metallicity GCs 
using different approaches. These studies include semi-analytical models that use galaxy assembly history from 
cosmological simulations and observed scaling relations to estimate the amount and metallicity of 
cold gas available for star formation~\citep{muratov2010}, physical models of the collapse leading 
to the formation of GCs based on high mass resolution simulations~\citep{griffen2010}, and
models based on the observed galaxy mass-metallicity relation, the 
galaxy stellar mass function and theoretical merger rates~\citep{tonini2013}. Other 
authors have explained the formation of low-metallicity GCs as starburst remnants of old dwarf
galaxies that could have entered the halos of spiral galaxies~\citep{elmegreen2012}. While these
works have explored multiple possible explanation of the peculiar observational traits of the GC 
populations, no agreement has yet been reached in the literature.

While the spectral and photometric properties of GC populations, as well as their radial distributions, 
have been explored in depth, relatively little work has addressed their two-dimensional (2D) 
distributions (e.g. NGC4471 in~\citealt{rhode2001}, NGC1399 in~\citealt{dirsch2003,bassino2006}, 
NGC3379, NGC4406 and NGC4594 in \citealt{rhode2004}, NGC4636 in~\citealt{dirsch2005}, multiple 
galaxies in~\citealt{hargis2012}, NGC3585 and NGC5812 in~\citealt{lane2013}). This type 
of work requires to extract the entire GC population with full or near-full coverage of the parent 
galaxy. Augmented by kinematics 
of the GC systems and subsystems (\citealt{strader2011} for M87;~\citealt{blom2012} for NGC4365), 
these studies are bringing forth a picture of complex and diverse GC populations in elliptical galaxies, 
consistent with a history of gravitational interactions and merging with neighboring galaxies.

Here we study the 2D GC distribution of NGC4261, for which deep observations of the innermost 
region of the galaxy, obtained with HST, are reported by~\cite{bonfini2012} (hereinafter B+12).
NGC4261 first came to attention because of the inhomogeneous spatial distribution 
of its low-mass X-ray binaries (LMXBs), which suggested a possible minor merger 
event~\citep{zezas2003}. Since a number of these LMXBs were associated with 
GCs~\citep{giordano2005}, the entire population of 718 GCs was extracted 
and studied with HST, revealing an azimuthal asymmetry in its distribution, which 
was attributed to past minor merging or interaction (B+12).~\cite{ferrarese1996}
also reported boxy isophotes which support the hypothesis of recent 
gravitational interaction experienced by NGC4261.

Although the results of B+12 are convincing, our more advanced analysis methods provide 
a clearer picture of the asymmetry and 
its statistical significance. Here we report the re-analysis of the 2D distribution of this GC population aimed at
quantifying the reality and shape  of the spatial features in the 2D GC distribution. We have used 
the K-Nearest Neighbor method 
(KNN,~\citealt{dressler1980}) to identify the 2D features, supplemented by Monte-Carlo simulations 
to test their statistical 
significance. In Section~\ref{sec:data} the data used in this paper are described. 
The method and the results of the its application 
to NGC4261 are discussed in Section~\ref{sec:method}. Our findings are discussed 
in Section~\ref{sec:discussion} and summarized in Section~\ref{sec:conclusions}.

\section{Data}
\label{sec:data}

\begin{figure*}[]
	\includegraphics[height=9cm,width=9cm,angle=0]{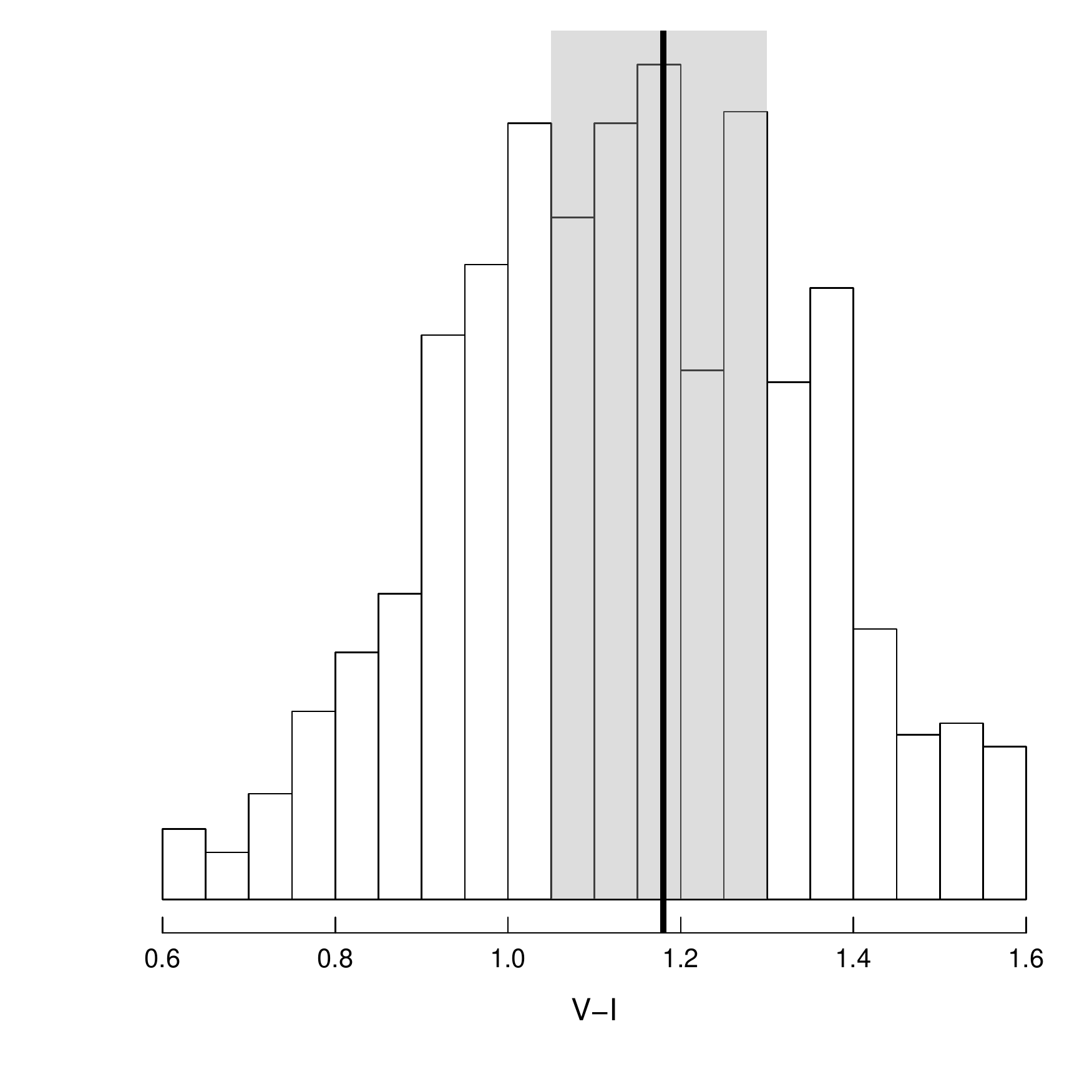}
	\includegraphics[height=9cm,width=9cm,angle=0]{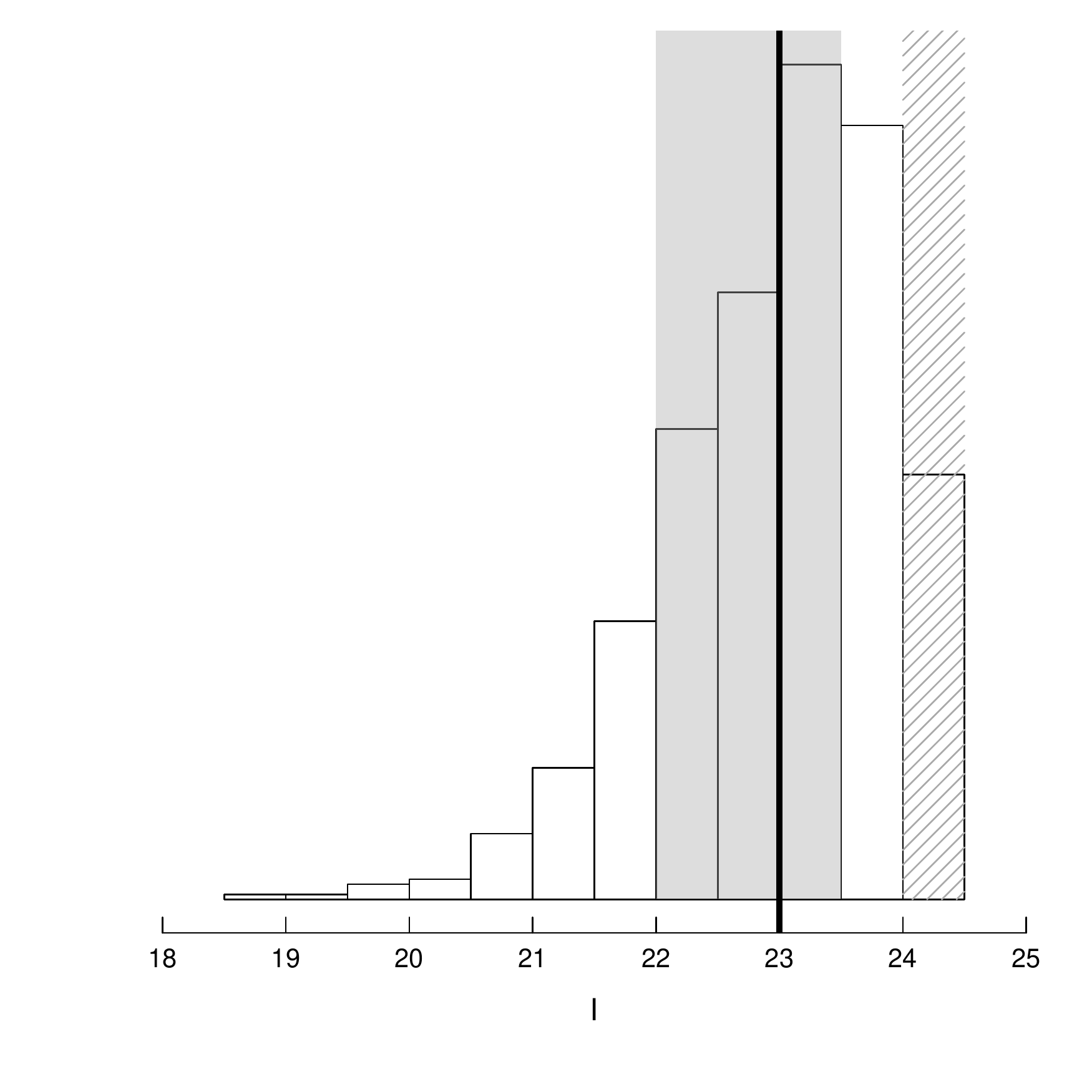}
	\caption{Left: Distribution of the $V\!-\!I$ color of the NGC4261 
	GCs sample. The vertical thick line ($V\!-\!I\!=\!1.15$) separates blue and red 
	GCs, and the shaded gray area indicates the interval of color thresholds 
	([1.05, 1.3]) which determine density and residual maps for the two color classes
	qualitatively similar to the maps obtained with the ($V\!-\!I\!=\!1.15$) color
	threshold (see discussion in Sec~\ref{sec:data} and Sec.~\ref{subsec:density}).
	Right: Distribution of the $I$ magnitude of the same sample. The vertical 
	thick line separates high luminosity and low luminosity GCs and the shaded
	gray area indicates the interval of $I$ magnitude value which determine density
	and residual maps similar to the maps obtained using the $I\!=\!23$ threshold
	(Sec.~\ref{subsec:density}). The shading lines region for $I\!>\!24$ has been not
	used to generate the density and residual maps for High-L and low-L GCs classes
	in Sec.~\ref{subsec:density}.}
       \label{fig:histocolori}
\end{figure*}

\begin{figure}[]
	\includegraphics[height=9cm,width=9cm,angle=0]{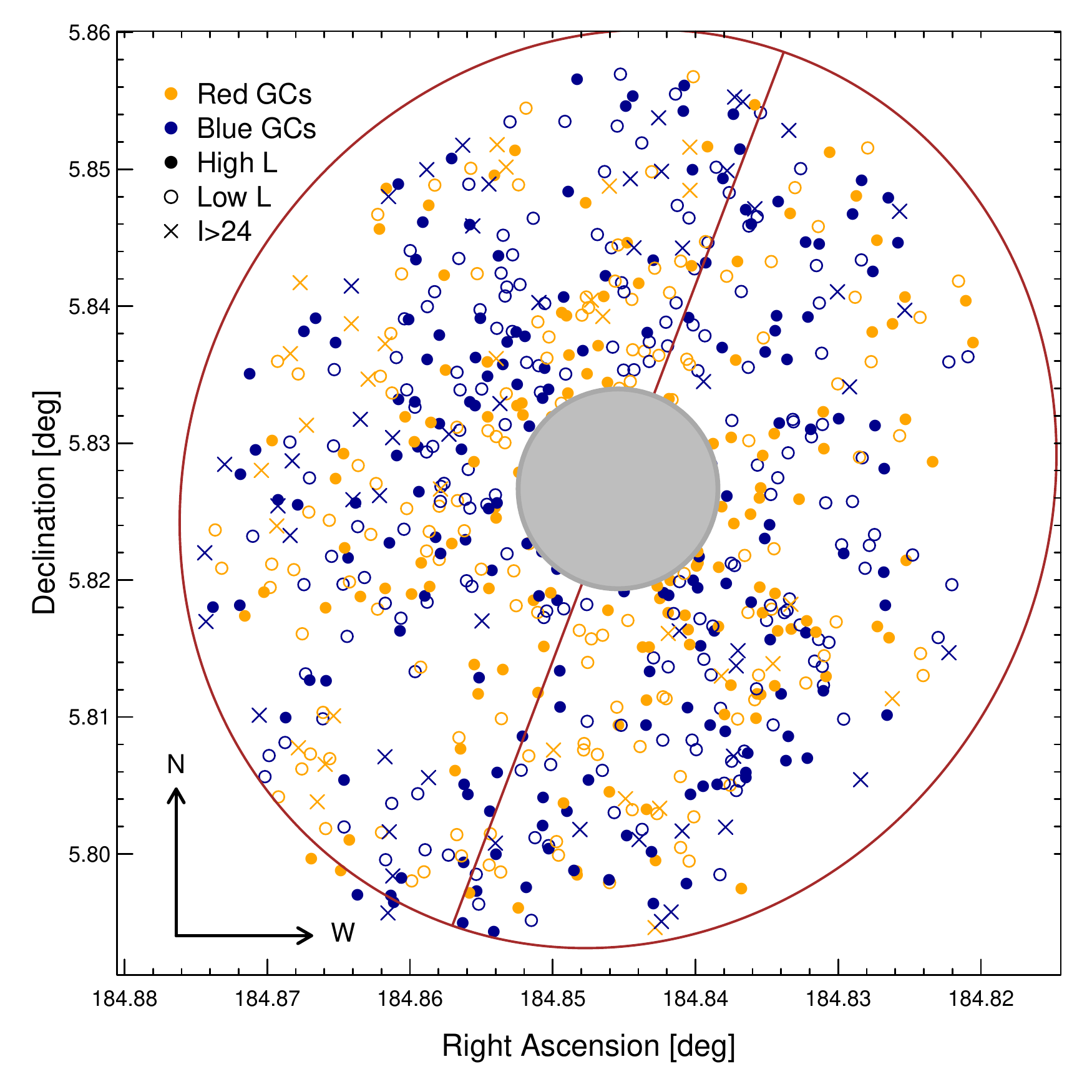}
	\caption{Positions of the GCs in NGC426. Blue and red GCs 
	are indicated as blue and orange symbols. High-luminosity ($I\!\leq\!23$) and 
	low-luminosity GCs ($I\!>\!23$) are indicated as solid and open symbols respectively, 
	while the crosses represent GCs fainter than $I\!\geq\!24$ that have been excluded
	from the luminosity classes analysis discussed in Section~\ref{subsec:density}.
	The central gray area has been excluded from our 
	analysis. The ellipse is the $D_{25}$ isophote from~\cite{devaucouleurs1991}.}
	\label{fig:positions}
\end{figure}

We have used the B+12 catalog of GC positions and properties, which lists
718 GCs, within the $D_{25}$ ellipse of NGC4261~\citep{devaucouleurs1991}. From this sample, 
we have extracted color and magnitude based subsamples. Figure~\ref{fig:histocolori} (left)
shows the $V\!-\!I$ histogram where blue ($V\!-\!I\!<\!1.15$) and red ($V\!-\!I\!\geq\!1.15$) GCs 
are separated 
following B+12. These authors found that $V\!-\!I\!=\!1.15$ is the color corresponding to 
the 50\% probability of the GC to belong to either the red
or the blue subpopulations, assuming a two-Gaussians model of the color distribution. However, 
as discussed in B+12, the GC color distribution does not show 
the clear bimodality typical of the GC color distribution in other early-type 
galaxies~\citep{brodie2006,peng2006}.
Figure~\ref{fig:histocolori} (right) shows the histogram of $I$ 
magnitudes; we arbitrarily define high luminosity (High L) GCs those with $I\!<\!23$ mag, 
and low luminosity (Low L) GCs those with $I\!\geq\!23$ mag. The $I\!<\!23$ value used to 
separate low-L from high-L GCs was set to obtain 
equipopulated classes of sources (see Tab.~\ref{tab:summary}). The density and 
residual maps obtained for the luminosity classes defined using $I\!<\!23$ are described in 
Sec.~\ref{subsec:density}.
Figure~\ref{fig:positions} shows the 
spatial distributions of GC positions in the plane of the sky, where the azimuthal 
asymmetry in the 2D projected GC distribution is evident. Table~\ref{tab:summary} 
lists the number of GCs in each of the main samples used in the following analysis. 
Changing the color and luminosity boundaries (shaded regions in the plots in Figure~\ref{fig:histocolori})
does not affect our results. Details can be found in Section~\ref{subsec:color}.

\begin{table}[h]
	\centering
	\caption{Summary of the samples of GCs observed in NGC4261 used in the paper.
	The number in parenthesis in the last column represent the GCs fainter than 
	$I\!=\!24$ which are not used to produce the density and residual maps for Low-L GCs.}
	\begin{tabular}{lccccc}
	\tableline
			& $N_{\mathrm{tot}}$	&$N_{\mathrm{red}}$ &$N_{\mathrm{blue}}$ 	& $N_{\mathrm{HighL}}$ & $N_{\mathrm{LowL}}$	\\			
	NGC4261	& 718				& 306			   & 412				& 316			        &  402(84)			\\
	\tableline
	\end{tabular}\\
	\label{tab:summary}
\end{table}

We have excluded from our analysis the central circular region with $r\!<\!0.42^{'}$ where 
incompleteness in the detection of the GCs is substantial (B+12). We have also 
excluded the undersampled regions outside the $D_{25}$ isophote~\citep{devaucouleurs1991}.

\begin{figure}[]
	\includegraphics[height=9cm,width=9cm,angle=0]{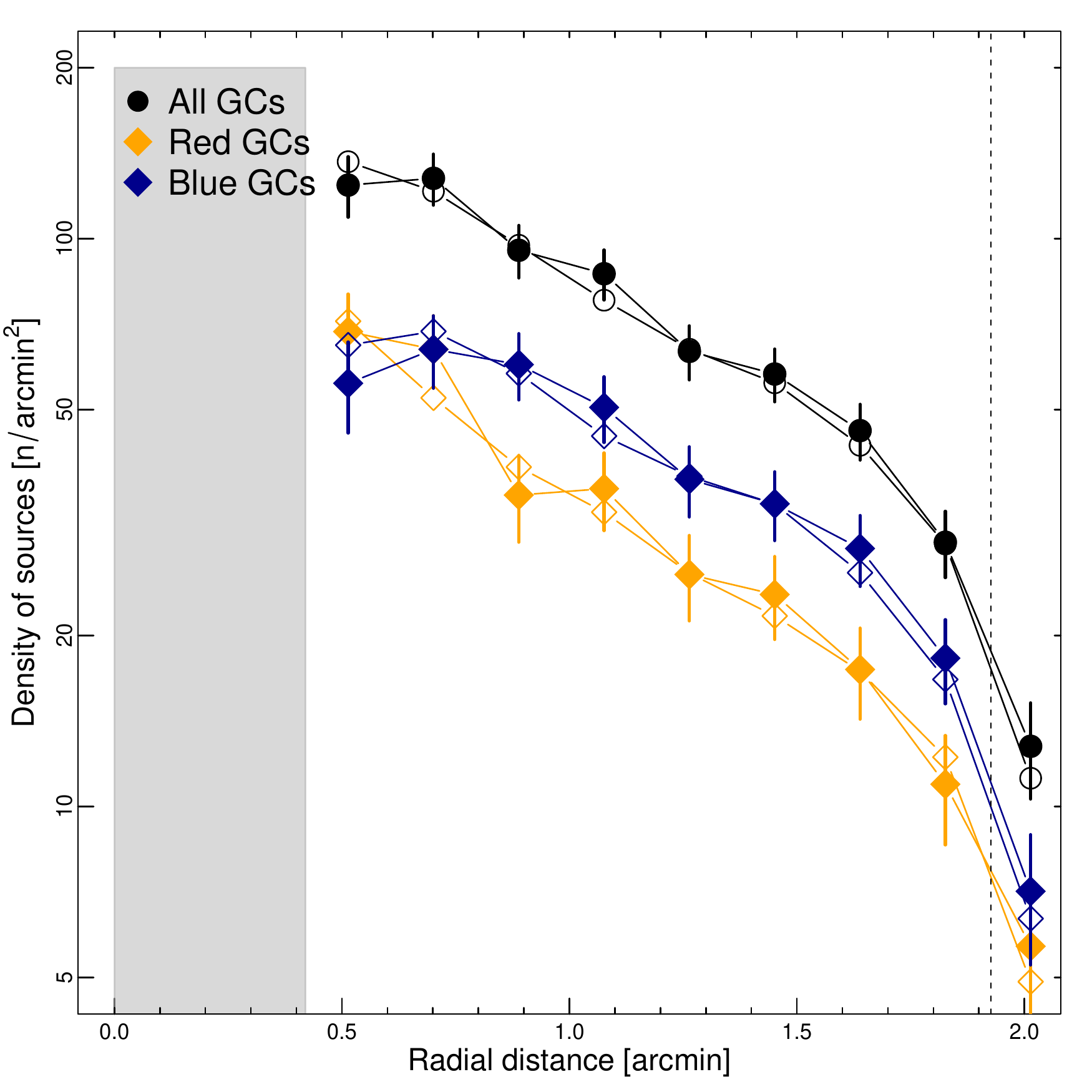}	
	\caption{Observed (solid) and simulated 
	(open) radial density profiles integrated in elliptical annuli for all (black symbols), red (orange
	symbols) and blue (blue symbols) GCs. The vertical dotted line marks the maximum radial distance of the 
	$D_{25}$ ellipse. The gray area has been excluded from our analysis.}
	\label{fig:radialdensity}
\end{figure}

\section{The Analysis}
\label{sec:method}

For each of the GCs samples listed in Table~\ref{tab:summary}, we have determined the
2D spatial distributions by applying the K-Nearest Neighbor (KNN) 
density estimator~\citep{dressler1980}. This density is based on the {\it local}
distribution of GCs, i.e. on the distances of the closest GCs. 
For each knot of a regular grid covering the region where the density is to be determined, 
we measured the distance of the $K$-th nearest 
neighbor GC ($D_{K}$) from the position of the knot. We used this approach in order to have density 
estimates even in the low density regions of the GC distribution. The point-density is estimated as: 

\begin{equation}
	D_{K}\!=\!\frac{K}{V_{D}(D_{K})} 
	\label{eq:knn}
\end{equation}

where K is the index of the nearest neighbor used to calculate the density; for example, 
for $K\!=\!5$ only the 5 GCs nearest to the grid knot are used. $V_{D}$ is the volume of the 
region within the distance $D_{K}$ of the $K$-th nearest neighbor from
the point where the density is evaluated. In the 
case of 2D spatial density, the volume $V_{D}\!=\!\pi\cdot d_{K}^2$ is equivalent to the area of the circle 
with radius equal to the 
distance of the $K$-th nearest neighbor. The assumption behind this method is that the density is
locally constant. The uncertainty on the KNN density scales with the square root of K, so that 
the relative fractional error is:

\begin{equation}
	\frac{\sigma_{D(K)}}{D(K)}\!=\!\frac{1}{\sqrt{K}}
\end{equation}

\noindent The fractional accuracy of the method increases with increasing K at the expense of 
the spatial resolution. 

For each subsample, we have determined the 2D GC surface density by using values of
$K$ ranging from 2 to 10. The density values have been calculated on a regular grid with 
spacing $\sim\!0.0018^{\circ}$ ($\sim6.5''$) and $\sim\!0.0015^{\circ}$ (5.4$''$) along the R.A.
and Dec. respectively, in order to have the same number of knots covering the whole region of 
the sky occupied by the GC distribution along the two coordinate axes. Different spacings along 
the R.A. and Dec. axes yield similar results to those described in the following. The density 
in the pixels overlapping the boundaries of the area covered by the observations has been 
weighted according to the fraction the area of the pixel located within the observed region.

To assess the significance of features suggested by the KNN density maps 
(hereafter ``observed'' density maps), we performed Monte-Carlo experiments, by creating 
random samples of GCs, each conforming to the observed radial density distribution of the 
relevant observed samples. The simulated random samples contain the same number 
of GCs as the corresponding observed samples.
The simulated radial positions were randomly drawn from the 
histogram of the observed radial distribution of GCs integrated over elliptical annuli with the 
same eccentricity and position angle of the $D_{25}$ isophote ($e\!=\!0.45$,~\cite{devaucouleurs1991}).
Independently, the 
azimuthal distributions were extracted from a uniformly random distribution between 0$^{\circ}$ 
and $360^{\circ}$. With this prescription, we do not need to determine a 2D
model of the observed GC distribution in NGC4261. We introduced a geometrical correction on the 
expected number of GCs to take into account the eccentricity of the galaxy. 

We also tried using a histogram generated from the best fit power law of the observed 
radial density profile to seed our simulations. This approach has the advantage of 
smoothing local density fluctuations and therefore, in principle, 
makes us more sensitive to small-scale annular features. However, it introduces
a model which may not represent the real radial 
distribution of GCs in NGC4261. Both approaches gave consistent results. In the following we 
will describe only the results obtained using the observed histogram to seed the simulations.

We performed 5000 simulations for each of the samples in Table~\ref{tab:summary} and constructed 
from each simulation the 2D density maps using the same KNN method applied to the observed data, 
for values of K ranging from 2 to 10. From these maps we derived average pixel-by-pixel simulated 
density maps for each sample under consideration (in the remainder of the paper, we will use 
``pixel'' to indicate each cell in the R.A. vs Dec. grid used to evaluate the 2D density associated to 
the GCs spatial distribution). Figure~\ref{fig:radialdensity} compares the 
observed and simulated mean radial density profiles for the color-based samples. 

Since our purpose is to characterize the 
deviations from the average radial trends of the GC distributions, our next step was to calculate the maps 
of residuals $R$ by subtracting pixel-by-pixel the mean simulated density map $<\!S\!>$ from the corresponding 
observed map $O$. These residual maps were also normalized  by the value of the density in 
the average simulated map $<\!S\!>$. The residual $R_{i}$ of the $i$-th pixel of the map is thus defined as:

\begin{equation}
	R_{i}\!=\frac{\!(O_{i}\!-\!<\!S\!>_{i})}{<\!S\!>_{i}}
	\label{eq:residuals}
\end{equation}

\begin{figure}[]
	\includegraphics[height=9cm,width=9cm,angle=0]{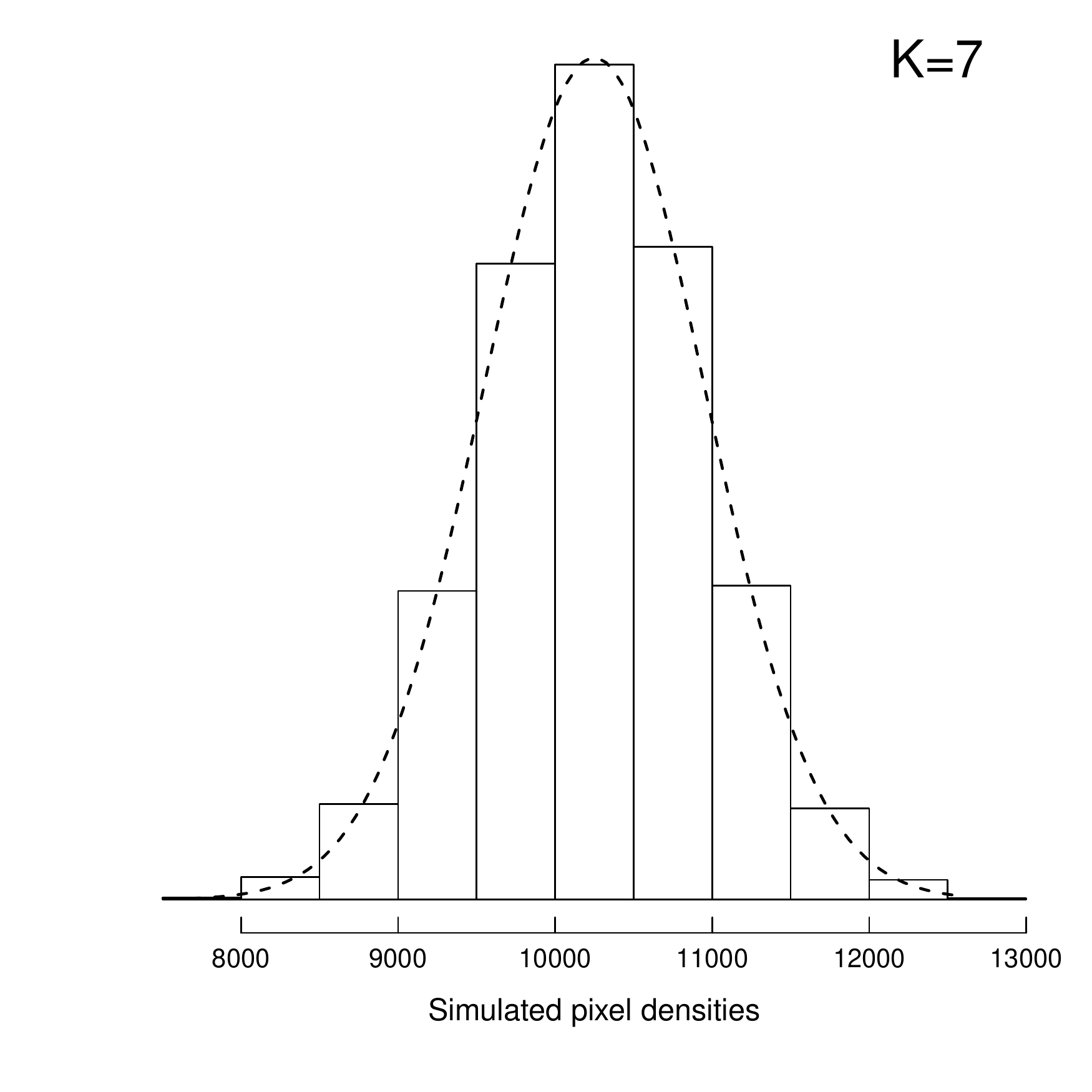}
	\caption{Representative histogram of the distributions of densities for a pixel of 
	the density maps derived from the 5000 simulated spatial 
	distributions of GCs for $K\!=\!7$. The dashed line is a the best-fit Gaussian model 
	of the histogram.}
       \label{fig:pixeldensitydistributions}
\end{figure}

We note that the pixel-by-pixel distributions of the simulated KNN densities are well approximated by 
Gaussians, simplifying the calculation of the statistical significance of the observed over densities. 
A typical example of the pixel density distributions is shown in Figure~\ref{fig:pixeldensitydistributions}.

To evaluate statistical significance of the observed residuals, for each set of simulated density 
maps we calculated the fraction of pixels with values above the 90-th percentile of the densities in the 
observed maps (the ``extreme'' pixels). 

Since the extreme pixels in the observed density and residual maps tend to be clearly 
spatially correlated, we also evaluated the fraction of simulations with at least one group of contiguous 
extreme pixels as large as the observed. For each simulation, we counted the number
of groups of extreme contiguous pixels with area (measured in pixels) equal to the area of the region
within the density contours corresponding to the 90-th percentile observed density threshold. 
Since we did not impose any specific geometry to the groups of contiguous
extreme pixels in the simulated density maps, these fractions represent upper limits to the 
fraction of extreme contiguous pixels expected for a given spatial distribution of residuals.

\begin{figure*}[]
	\includegraphics[height=9cm,width=9cm,angle=0]{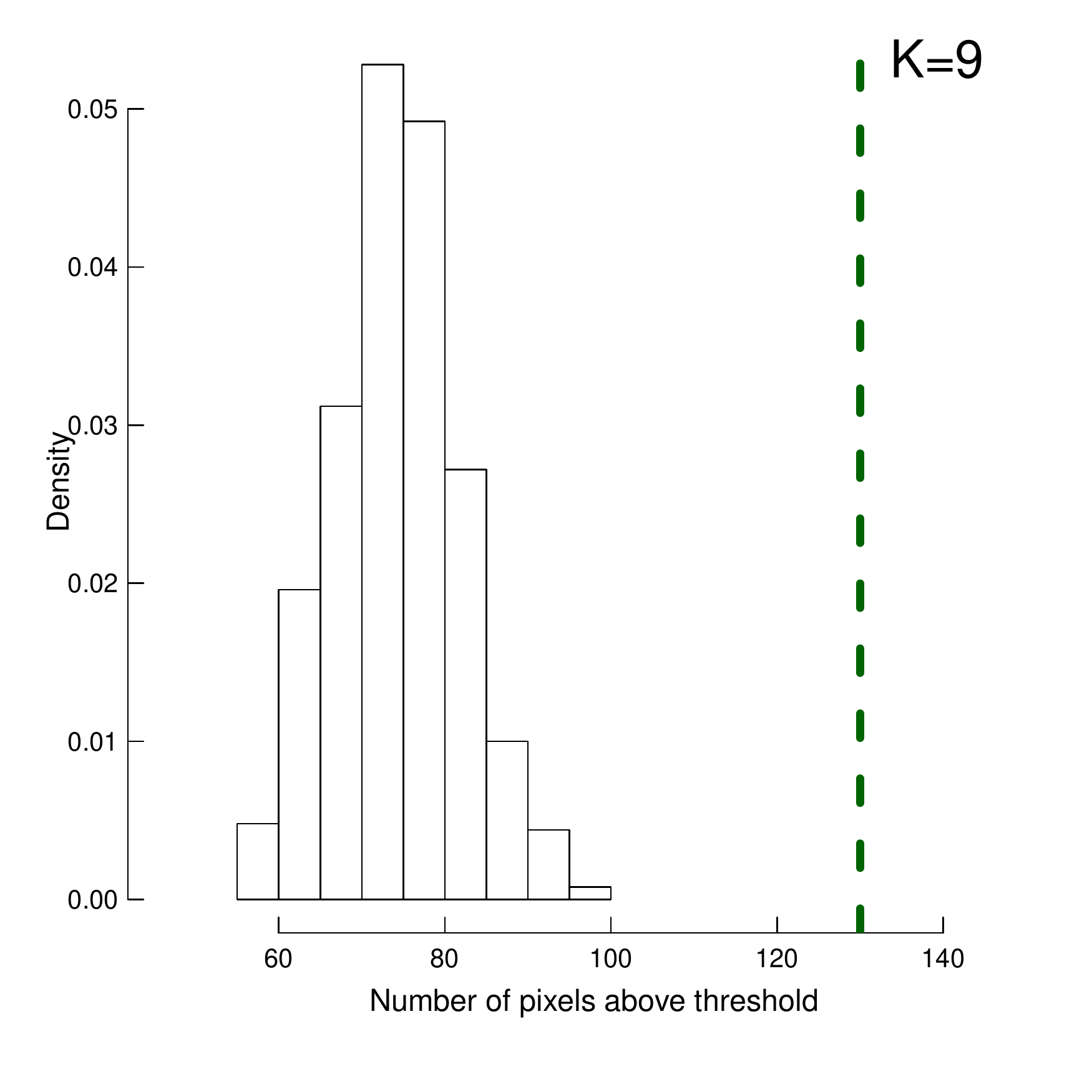}
	\includegraphics[height=9cm,width=9cm,angle=0]{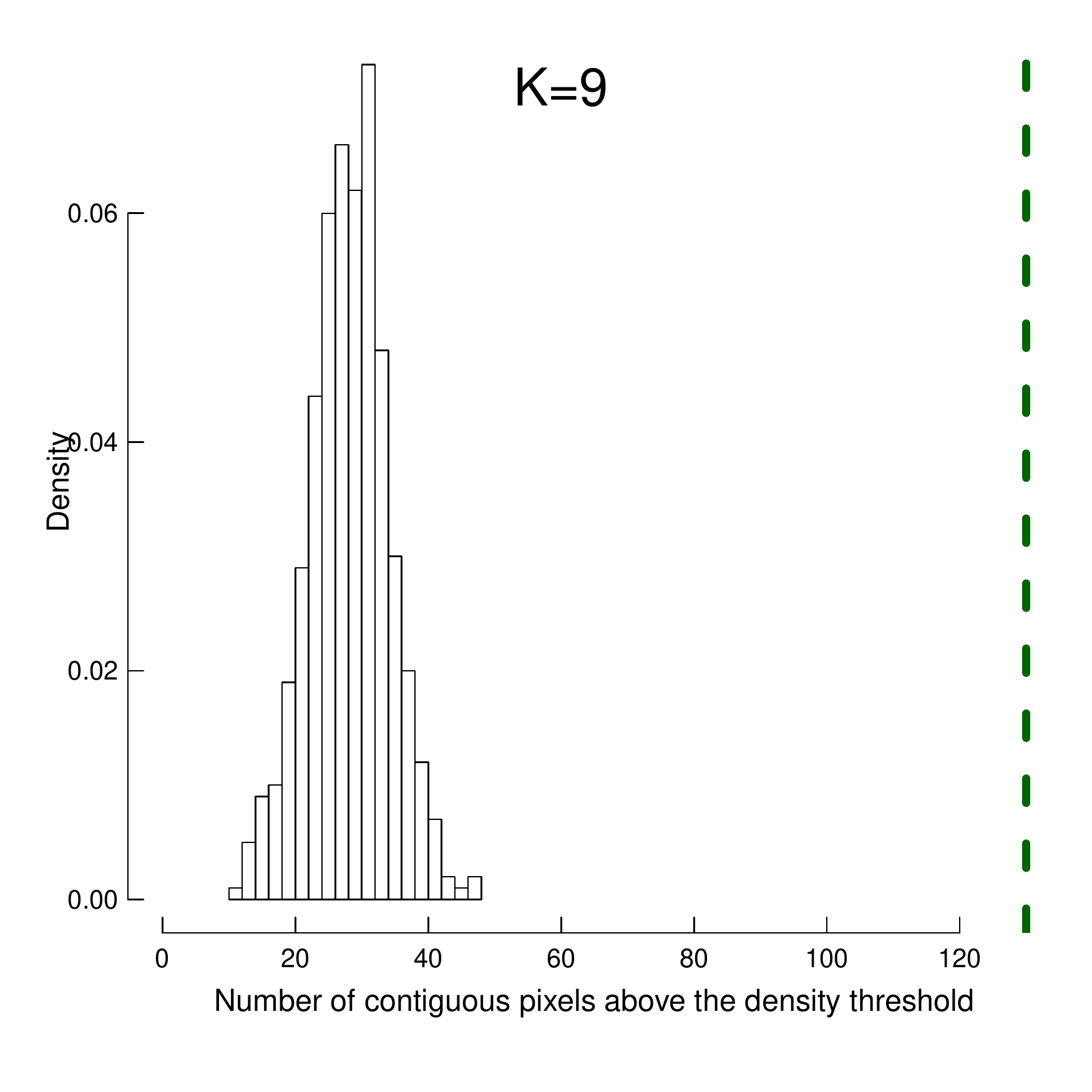}	
	\caption{Left: histograms of the number of pixels in the simulated two-dimensional GCs 
	density maps of the entire GC sample (calculated for $K\!=\!9$) with density $>$ 90-th percentile of the 
	distribution of pixels in the observed density map. Right: same as above, with the additional 
	constraint of spatial 
	correlation as strong as the high-density pixels in the observed maps. In both plots, the vertical lines 
	represent the number of observed pixels whose density exceeds the density thresholds.}
	\label{fig:extremepixels}
\end{figure*}

The results, compiled in Table~\ref{tab:statistics}, show 
that for small values of K, simulated density maps have a total fraction of extreme pixels larger 
than the observed. However, this fraction decreases with increasing K, and it becomes negligible for 
$K\!\geq\!5$. The percentage of contiguous extreme pixels (in parenthesis in 
Table~\ref{tab:statistics}) does not exceed 20\% even for $K\!=\!2$ and rapidly 
decreases to zero at $K\!=\!5$. 
Therefore, the probability of a random distribution of high-density pixels with the observed spatial 
distribution is very low even for low $K$ and becomes null with increasing $K$. Based on 
Table~\ref{tab:statistics}, in the following we will only discuss our results for $K\!\geq\!5$. 
Figure~\ref{fig:extremepixels} shows the histogram of the simulated number 
of extreme pixels compared to the observed for $K\!=\!9$ for the entire sample, and the corresponding 
histograms for spatially correlated extreme pixels. Together with Table~\ref{tab:statistics}, these 
comparisons shows that the degree of correlated features observed (discussed in detail below) 
in the data cannot be randomly produced.

\begin{table*}
	\centering
	\caption{Fractions of simulated density maps 
	with number of extreme pixels (i.e., pixels with density values exceeding the 90-th 
	percentile of the observed pixel density 
	distribution) larger than the number of observed extreme pixels. Values in parenthesis 
	refer to the fraction of simulated density maps with at least one 
	group of contiguous extreme pixels as large as the groups of contiguous extreme pixels
	in the observed density maps (see details in Section~\ref{sec:method}).
	These fractions were determined by counting the number of simulated density
	maps with at least one group of contiguous extreme pixels equal or larger than
	the group of contiguous extreme pixels observed over-density regions.}
	\begin{tabular}{lcccccccccc}
	\tableline
	NGC4261 (density)		& 				&				&				&			&			&			&			&			&			\\
		 				&$K\!=\!2$			&$K\!=\!3$			&$K\!=\!4$			&$K\!=\!5$		&$K\!=\!6$		&$K\!=\!7$		&$K\!=\!8$		&$K\!=\!9$		&$K\!=\!10$	\\
	\tableline
	All GCs (red$+$blue)	&100\%(12.4\%)	&100\%(6.6\%)		&55.2\%(0.9\%)	&0.2\%(0\%)	&0\%(0\%)	&0\%(0\%)	&0\%(0\%)	&0\%(0\%)	&0\%(0\%)	\\
	Red GCs				&100\%(19.9\%)	&99.8\%(12\%)		&20.8\%(2.1\%)	&0.2\%(0\%)	&0\%(0\%)	&0\%(0\%)	&0\%(0\%)	&0\%(0\%)	&0\%(0\%)	\\
	Blue GCs				&100\%(17.3\%)	&100\%(9.8\%)		&77.4\%(1.3\%)	&3.2\%(0\%)	&0\%(0\%)	&0\%(0\%)	&0\%(0\%)	&0\%(0\%)	&0\%(0\%)	\\	
	\tableline
	NGC4261 (residual)		& 				&				&				&			&			&			&			&			&			\\
		 				&$K\!=\!2$			&$K\!=\!3$			&$K\!=\!4$			&$K\!=\!5$		&$K\!=\!6$		&$K\!=\!7$		&$K\!=\!8$		&$K\!=\!9$		&$K\!=\!10$	\\
	\tableline
	All GCs (red$+$blue)	&100\%(4.4\%)		&100\%(0.5\%)		&55.2\%(0.1\%)	&0.2\%(0\%)	&0\%(0\%)	&0\%(0\%)	&0\%(0\%)	&0\%(0\%)	&0\%(0\%)	\\
	Red GCs				&100\%(3.9\%)		&99.8\%(1.4\%)	&20.8\%(0\%)		&0.2\%(0\%)	&0\%(0\%)	&0\%(0\%)	&0\%(0\%)	&0\%(0\%)	&0\%(0\%)	\\
	Blue GCs				&100\%(1.3\%)		&100\%(0.2\%)		&77.4\%(0.2\%)	&3.2\%(0\%)	&0\%(0\%)	&0\%(0\%)	&0\%(0\%)	&0\%(0\%)	&0\%(0\%)	\\	
	\tableline
	\end{tabular}\\
	\label{tab:statistics}
\end{table*}

\subsection{Density and Residual maps}
\label{subsec:density}

\begin{figure*}[]
	\includegraphics[height=6cm,width=6cm,angle=0]{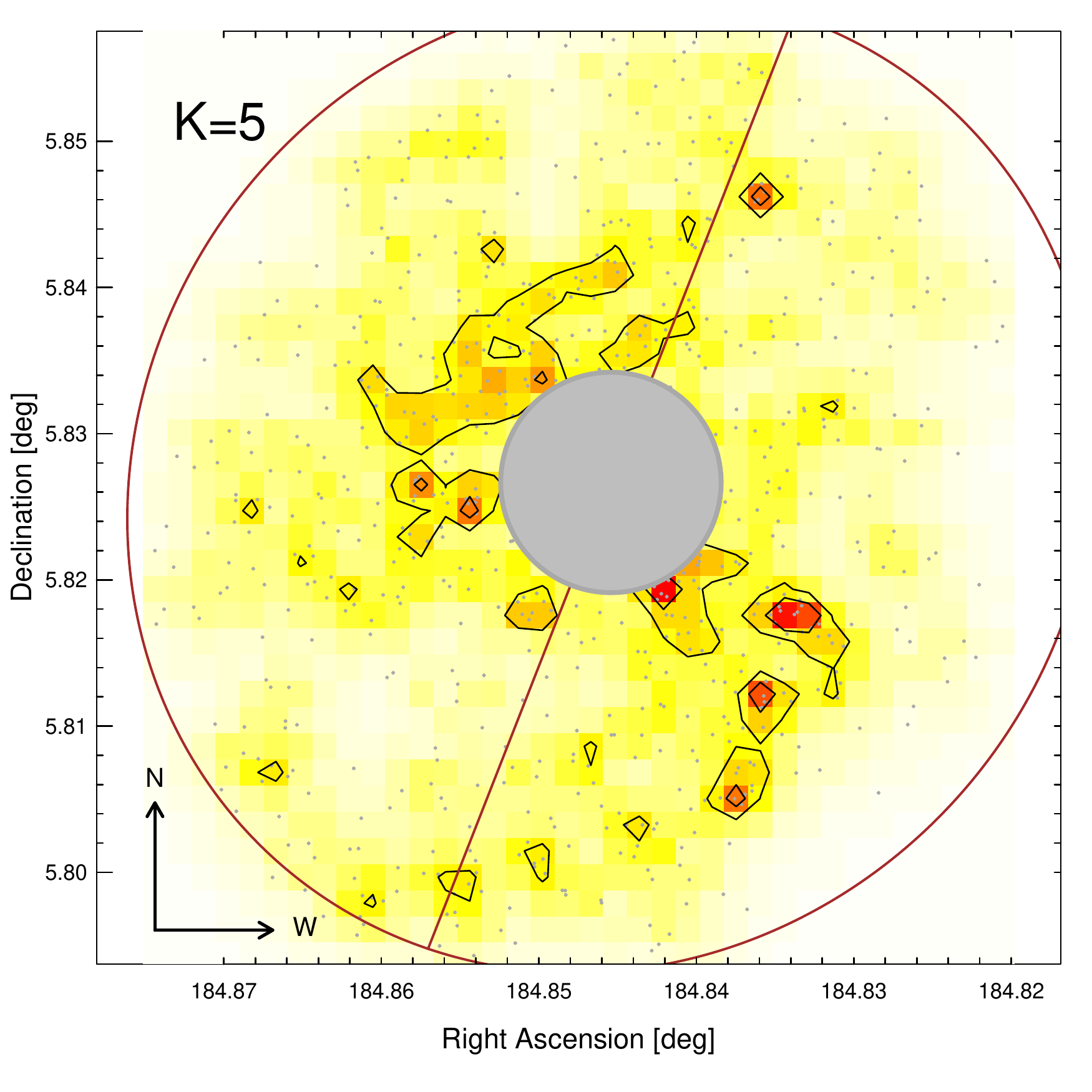}
	\includegraphics[height=6cm,width=6cm,angle=0]{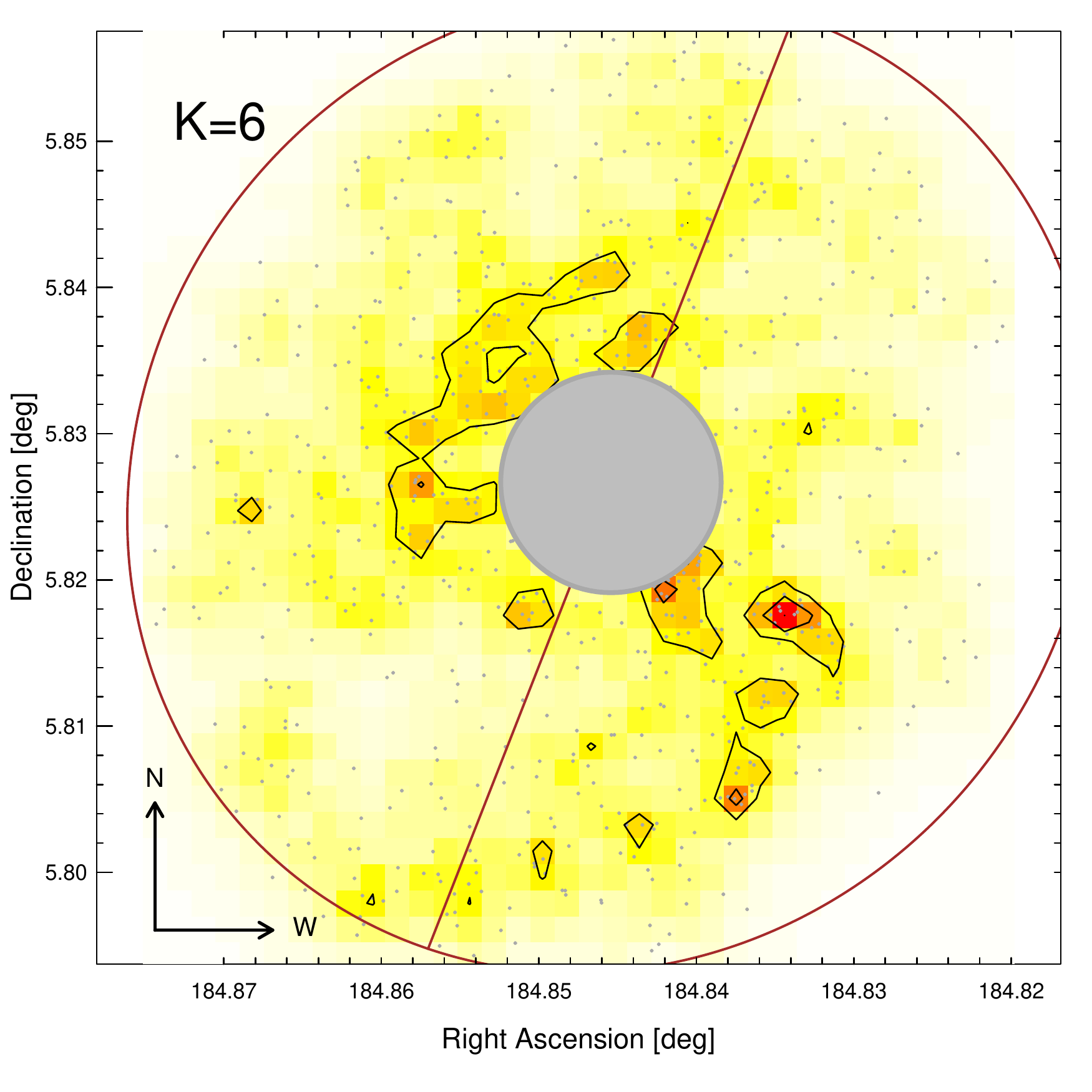}	
	\includegraphics[height=6cm,width=6cm,angle=0]{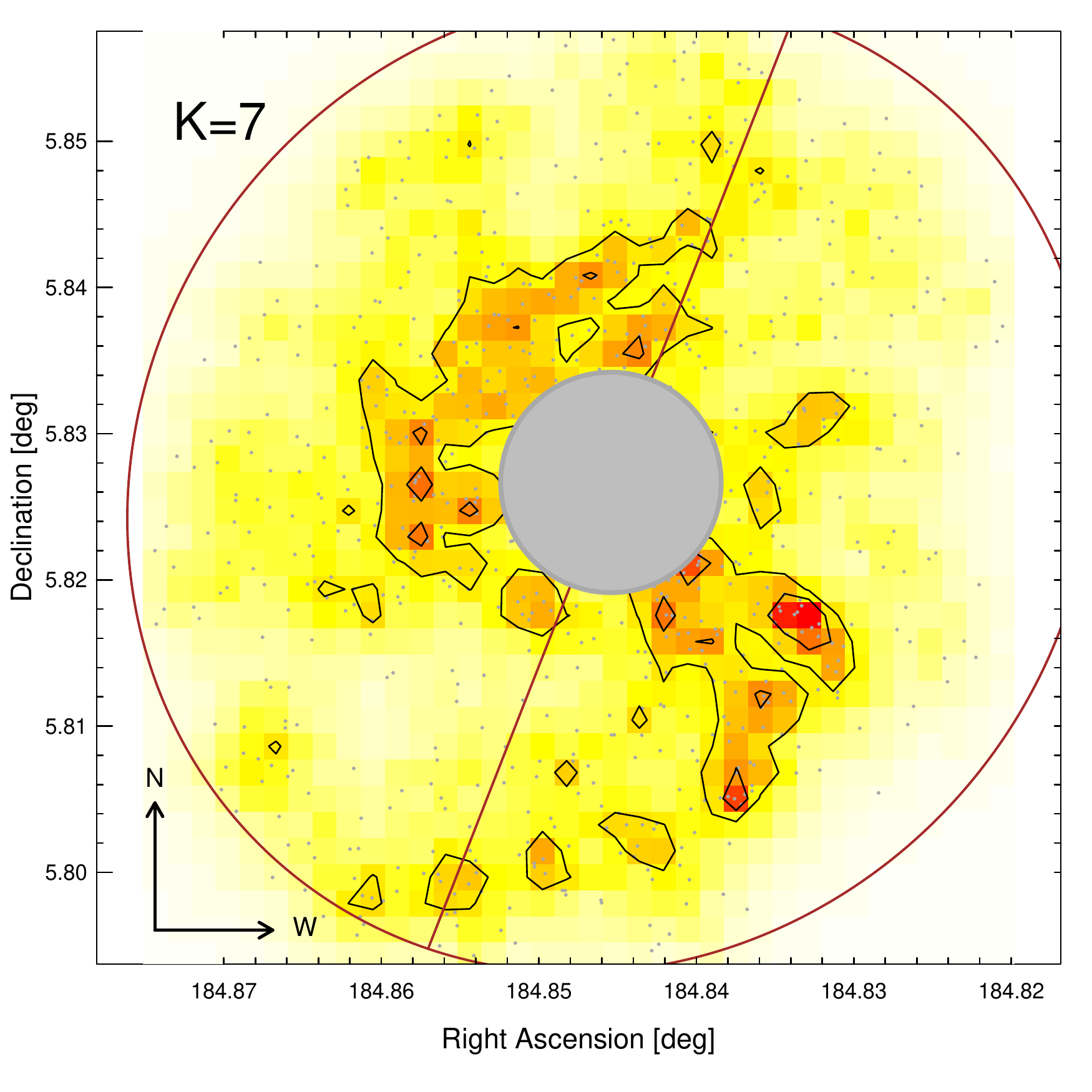}\\
	\includegraphics[height=6cm,width=6cm,angle=0]{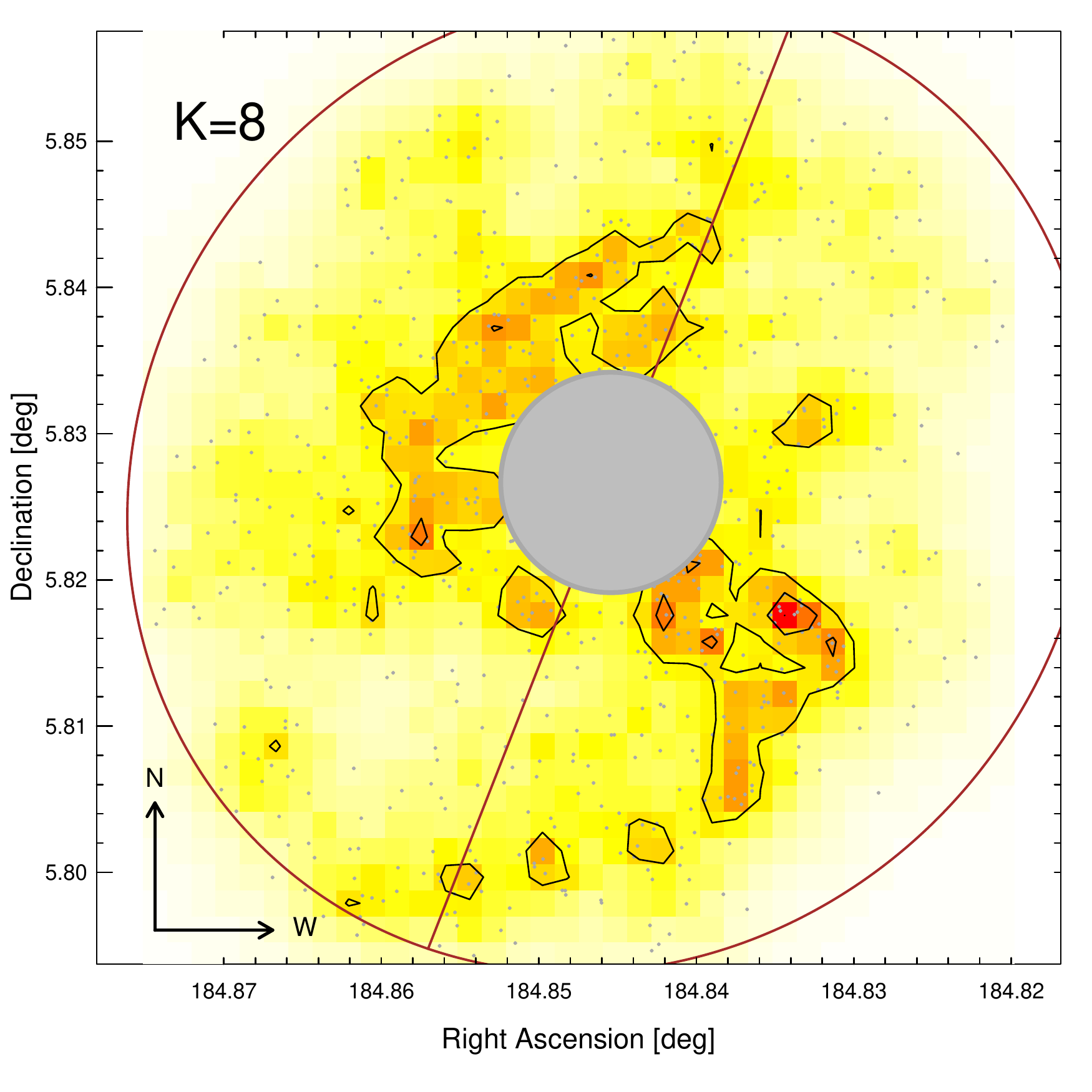}
	\includegraphics[height=6cm,width=6cm,angle=0]{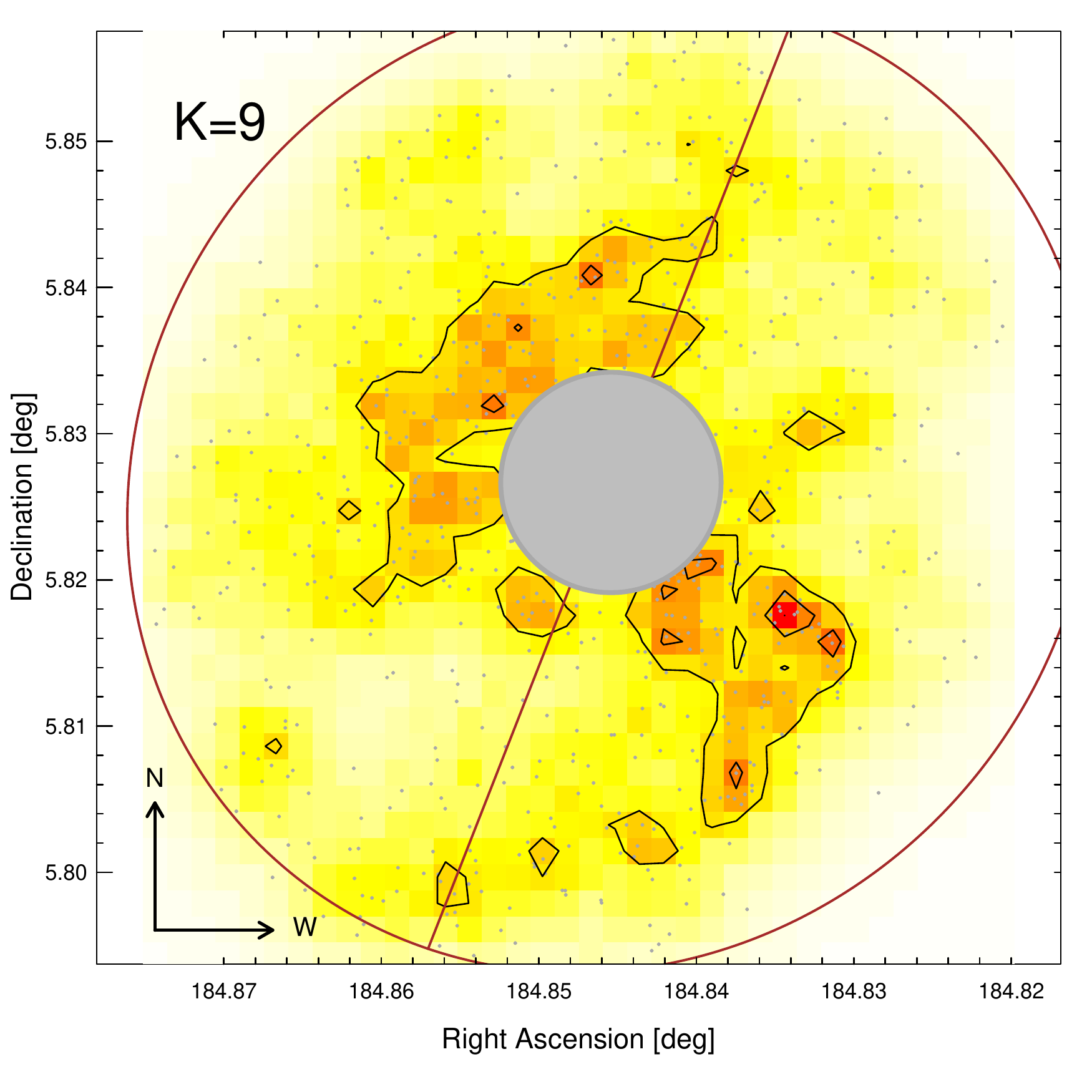}	
	\includegraphics[height=6cm,width=6cm,angle=0]{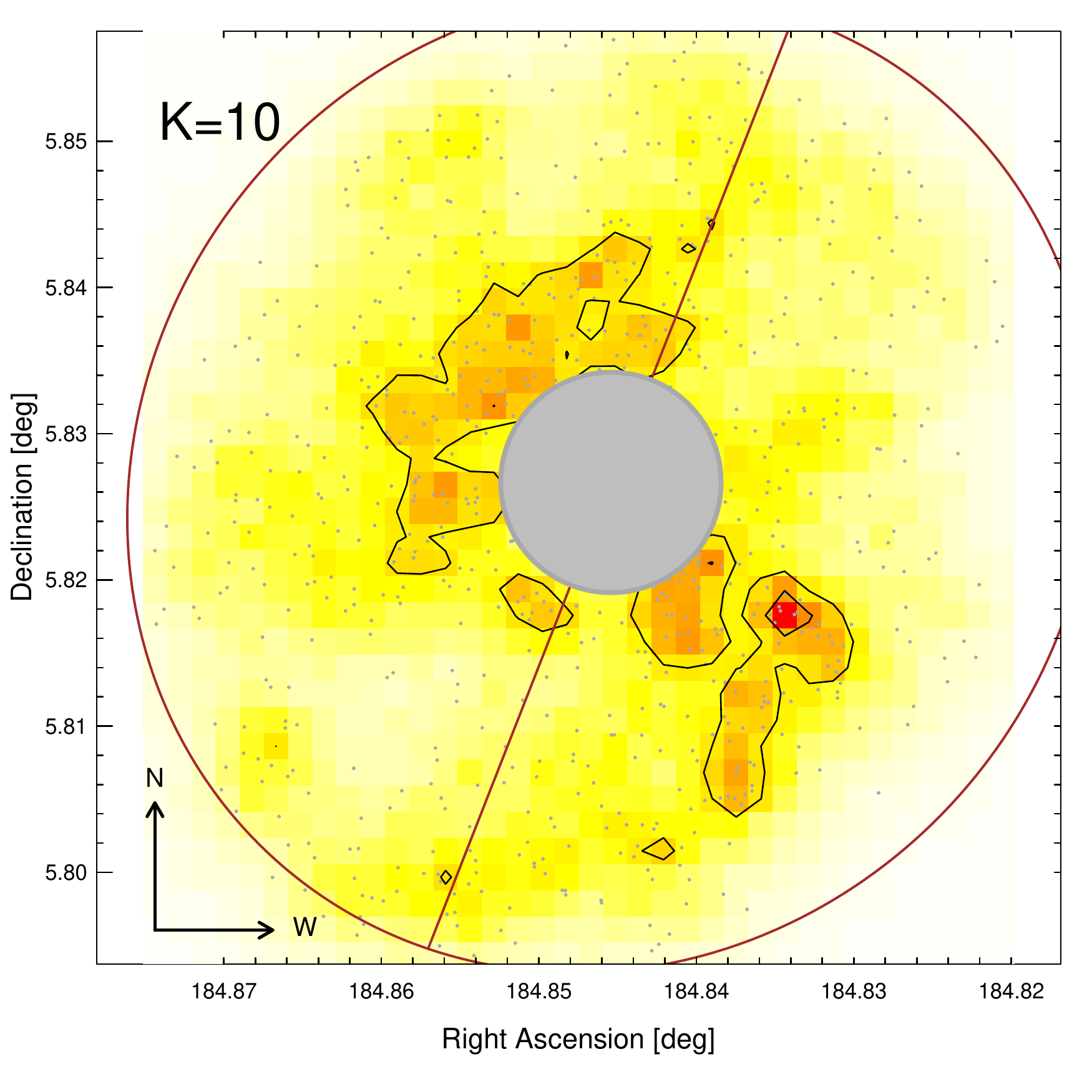}\\
	\caption{KNN density maps of the entire GC sample of NGC4261. Arbitrary isodensity contours 
	show the higher-density regions in each map.}
	\label{fig:2dmapsngc4261}
\end{figure*}

Figure~\ref{fig:2dmapsngc4261} shows the 2D KNN density maps for the spatial distribution of the entire 
GC sample for $K$ ranging from 5 to 10. In all these maps two high-density structures emerge in the 
S-W and N-E quadrants, on opposite sides of the galaxy 
major axis. The S-W density enhancement resembles a spiral arm and is composed of two distinct sections: a short 
elongated structure aligned along the radial direction and a second longer feature spanning azimuthally more 
than 45$^{\circ}$. The N-E high-density region has an elongated shape and occupies azimuthally 
$\sim\!100^{\circ}$. These over-densities roughly correspond to the local enhancements of the GC 
distribution shown in 
Figure 9 of the B+12 paper, which was obtained by applying adaptive kernel smoothing and 
Voronoi Percolation-Tessellation. This figure shows that maps with $K\!=\!\{7, 8, 9\}$ are similar, while the 
$K\!=\!10$ maps smooth over some of the spatial features. In the remainder of the paper, we will discuss 
only the density and residual maps obtained for $K\!=\!9$ because these maps are the most statistically 
significant (as discussed in Section~\ref{sec:method} and shown in Table~\ref{tab:statistics}).

\begin{figure}[]
	\includegraphics[height=9cm,width=9cm,angle=0]{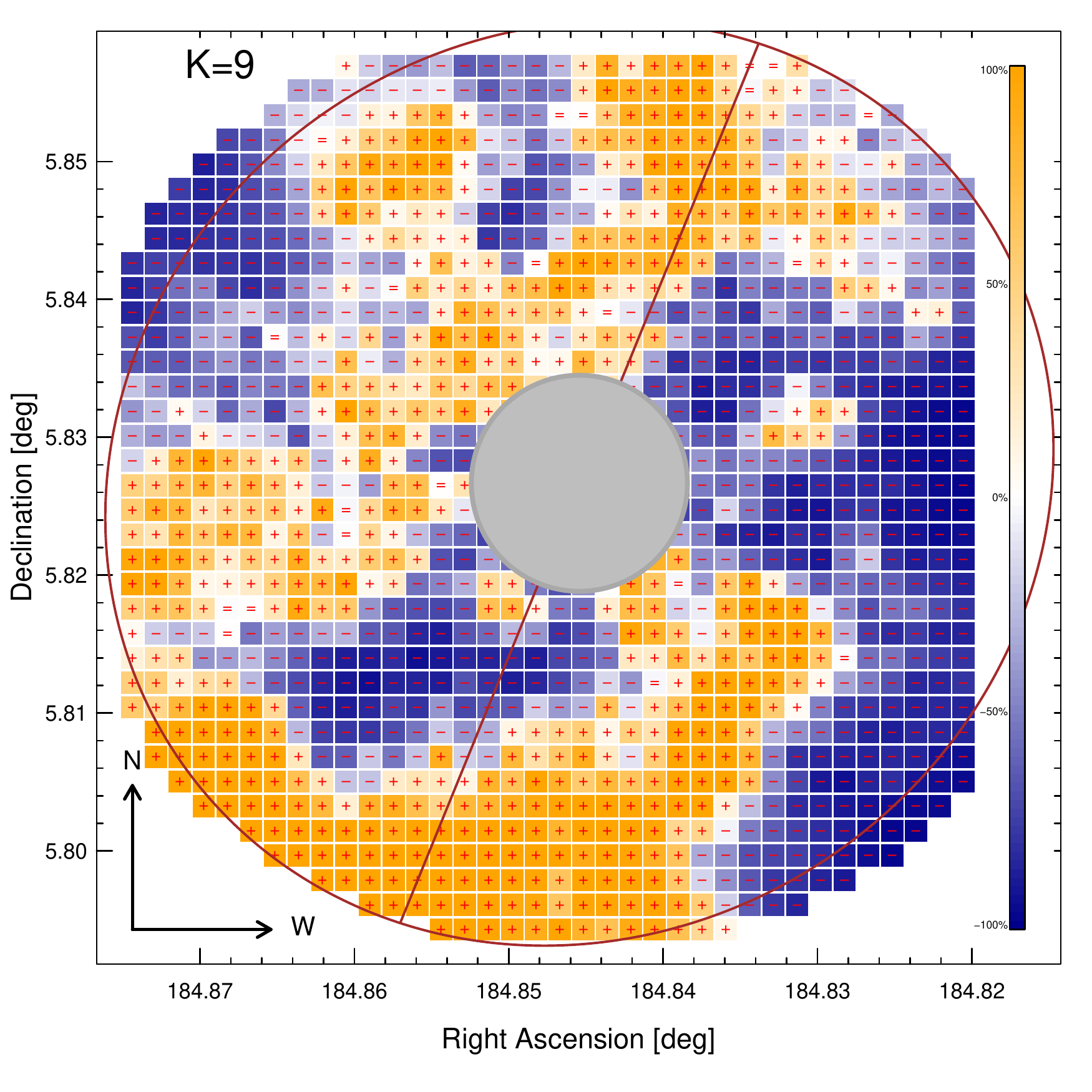}
	\caption{$K\!=\!9$ residual map 
	of the entire GCs sample. Pixels are color-coded 
	according to the number of $\sigma$ the pixel deviates from the average. Darker colors indicate larger
	residuals: blue, negative; orange, positive. The small $+, -$ and $=$ signs within each 
	pixel indicate positive, negative or null residuals respectively.}
	\label{fig:res2dmapsngc4261}
\end{figure}

\begin{figure*}[]
	\includegraphics[height=6cm,width=6cm,angle=0]{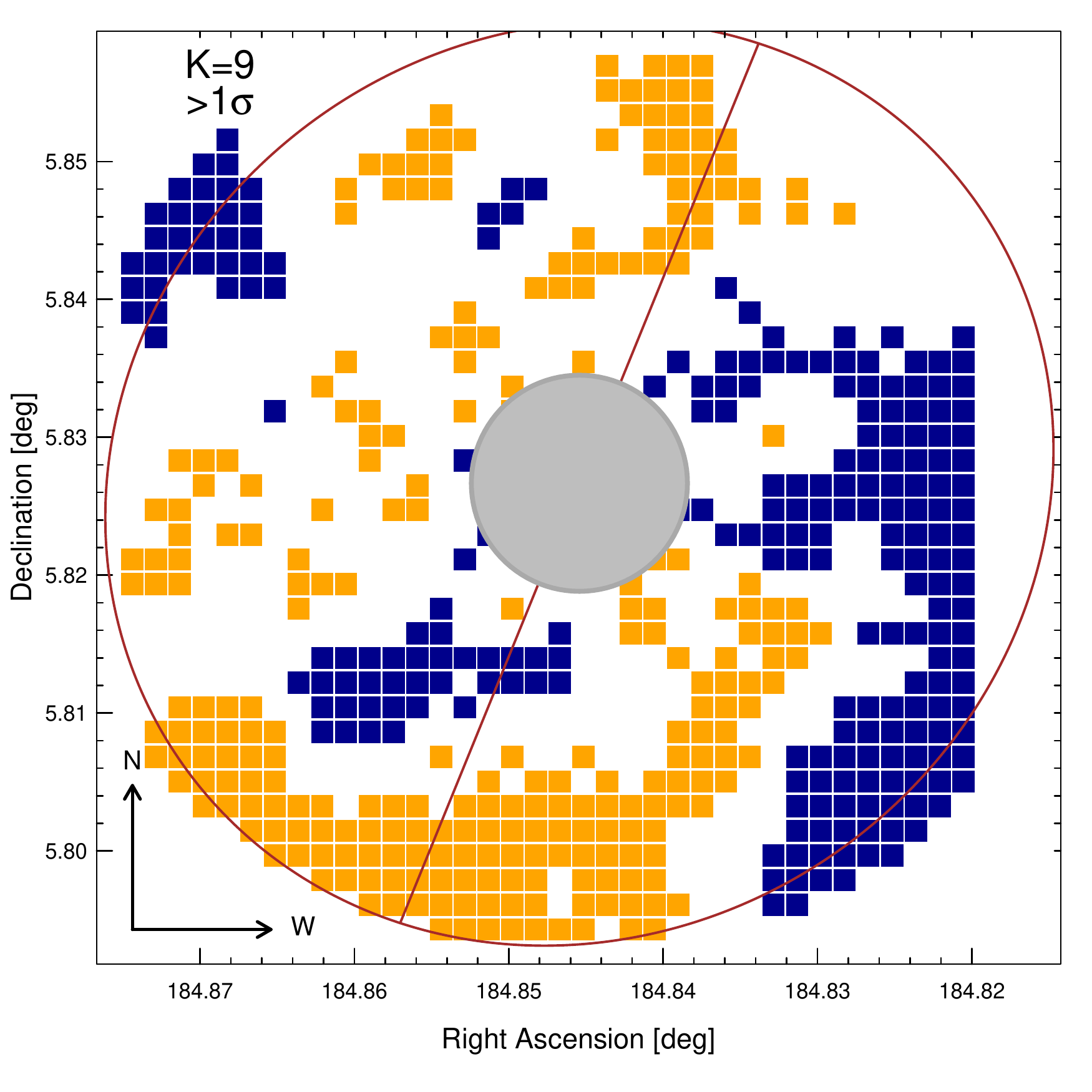}
	\includegraphics[height=6cm,width=6cm,angle=0]{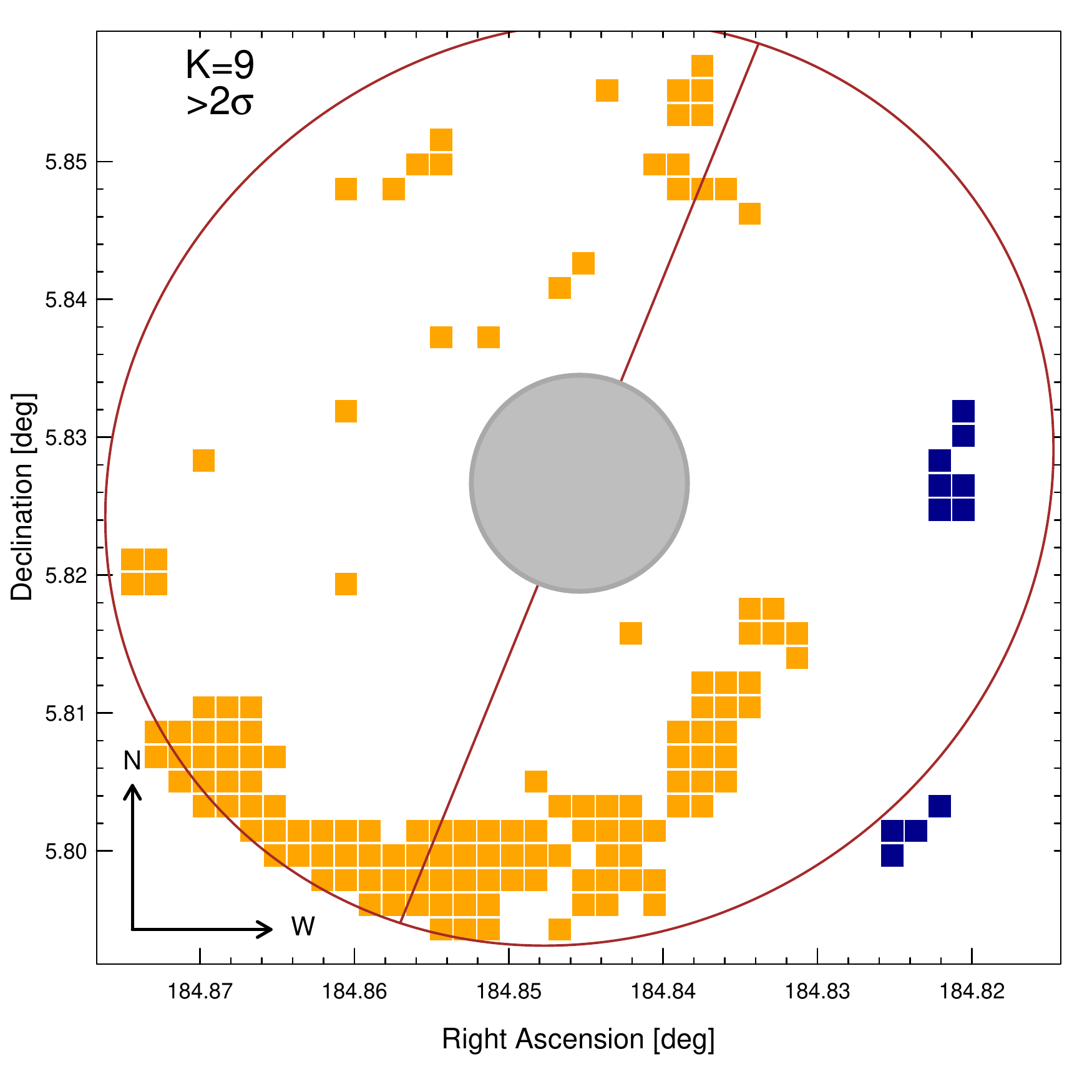}	
	\includegraphics[height=6cm,width=6cm,angle=0]{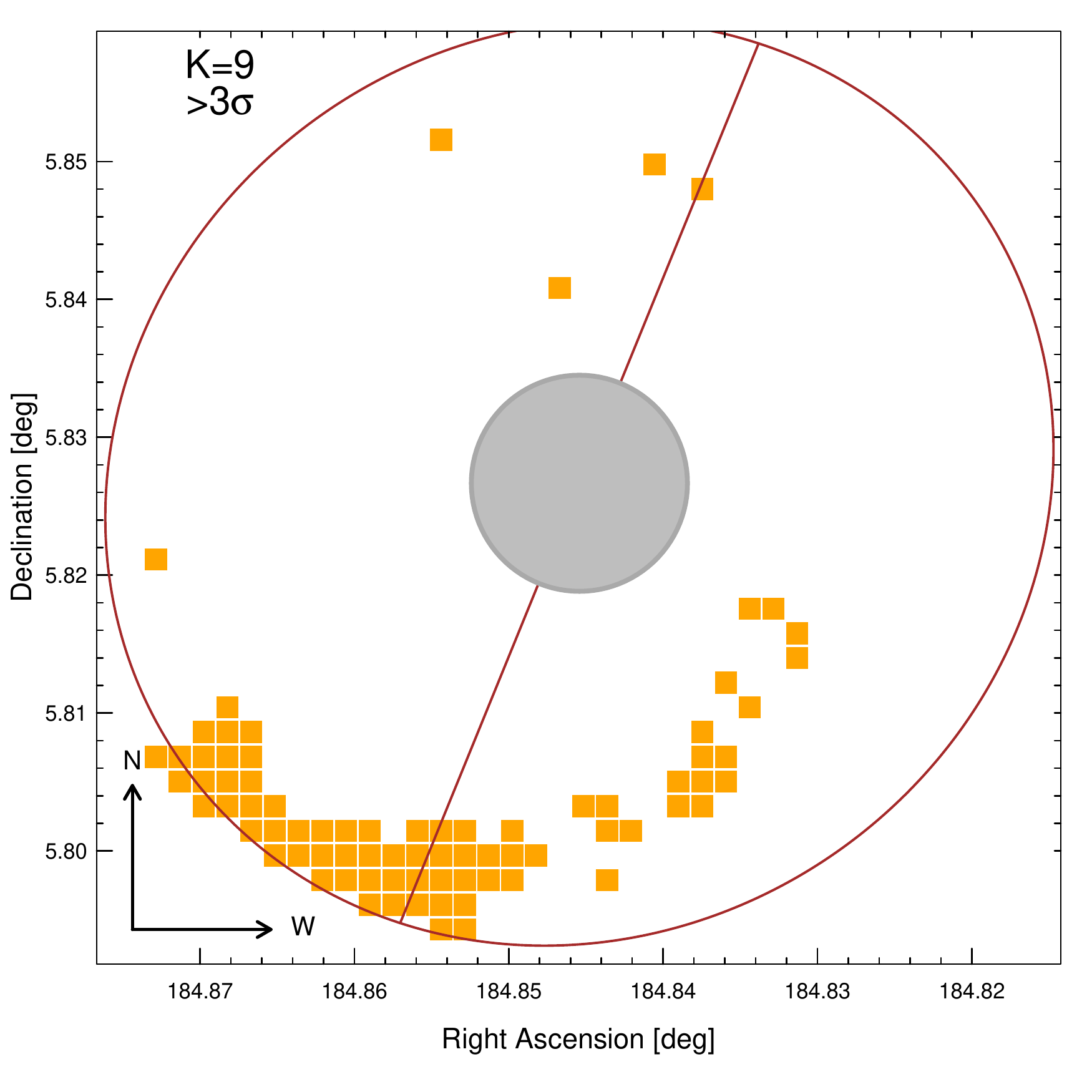}		
	\caption{From left to right, positions of the $K\!=\!9$ residuals with significance larger than 
	1$\sigma$, 2$\sigma$ and 3$\sigma$.}
       \label{fig:res2dmapsngc4261sigmas}
\end{figure*}

We show
in Figure~\ref{fig:res2dmapsngc4261} the $K\!=\!9$ residual map, which highlights the existence of the two 
main over-densities in the S-E and N-W quadrants already observed in the density maps and discussed 
above, together
with similarly significant under-densities in the S-E region and the western region of the galaxy. The latter is 
partially due to the incomplete HST coverage, see Figure 1 in B+12 for details.
Figure~\ref{fig:res2dmapsngc4261sigmas} maps the single pixels with significance larger than 
1$\sigma$, 2$\sigma$ and 3$\sigma$. In particular the S-W features
stands out, since multiple spatially correlated nearby pixels with significant positive residuals are located in this 
region. Although the N-E over-density is weaker, the spatial correlation of the pixels with 
positive residuals in this case cannot be due to chance, as shown in Table~\ref{tab:statistics}. 
The number of GCs located within the pixels associated to positive residuals is 175 ($\sim\!24\%$), 
86 ($\sim\!12\%$) and 60 ($\sim\!8\%$) for 1$\sigma$, 2$\sigma$ and 3$\sigma$ respectively. 
Table~\ref{tab:numgcsover} shows the fractions of GCs located within the statistically significant 
over-densities for all GCs classes used in this paper.

\begin{table*}
	\centering
	\caption{Number of GCs located in over-densities regions with significance larger than 1$\sigma$, 
	2$\sigma$ and 3$\sigma$ respectively for all classes of GCs used in this paper. In parenthesis, the 
	percentage relative to the total number of sources in each class, as shown
	in Tab.~\ref{tab:summary}.}
	\begin{tabular}{lccccccccc}
	\tableline
			& \multicolumn{3}{ |c| }{All GCs} & \multicolumn{3}{ |c| }{Red GCs} & \multicolumn{3}{ |c| }{Blue GCs}\\
			& $1\sigma$&$2\sigma$&$3\sigma$&$1\sigma$&$2\sigma$&$3\sigma$&$1\sigma$&$2\sigma$&$3\sigma$\\
	\tableline
	$K\!=\!5$	&241($33.5\%$)&137($19.1\%$)&78($11.0\%$)&203($66.3\%$)&109($35.6\%$)&79($25.8\%$)&191($46.4\%$)&100($24.3\%$)&67($16.3\%$)\\
	$K\!=\!6$	&241($33.5\%$)&133($18.5\%$)&80($11.1\%$)&217($70.9\%$)&109($35.6\%$)&79($25.8\%$)&219($53.2\%$)&117($28.4\%$)&60($14.6\%$)\\
	$K\!=\!7$	&247($34.4\%$)&125($17.4\%$)&88($12.3\%$)&210($68.6\%$)&122($38.9\%$)&80($26.1\%$)&228($55.3\%$)&125($30.3\%$)&70($17.0\%$)\\
	$K\!=\!8$	&247($34.4\%$)&133($18.5\%$)&79($11.0\%$)&218($71.2\%$)&141($46.1\%$)&93($30.4\%$)&225($54.6\%$)&112($27.2\%$)&76($18.4\%$) \\
	$K\!=\!9$	&252($35.1\%$)&137($19.1\%$)&90($12.5\%$)&211($70.0\%$)&128($41.8\%$)&91($29.7\%$)&230($55.8\%$)&126($30.6\%$)&63($15.3\%$) \\
	$K\!=\!10$	&274($38.1\%$)&140($19.5\%$)&93($13.0\%$)&220($71.9\%$)&135($44.1\%$)&78($25.5\%$)&224($54.4\%$)&143($34.7\%$)&77($18.7\%$) \\				
	\tableline
			&  \multicolumn{3}{ |c| }{High-L GCs} & \multicolumn{3}{ |c| }{Low-L GCs}\\
			& $1\sigma$&$2\sigma$&$3\sigma$&$1\sigma$&$2\sigma$&$3\sigma$\\

	\tableline
	$K\!=\!5$	&194($61.4\%$) &104($33.0\%$)&58($18.3\%$)&200($62.9\%$) &103($32.4\%$)	&73($23.0\%$)		\\
	$K\!=\!6$	&210($66.5\%$) &104($33.0\%$)&63($19.9\%$)&208($65.4\%$) &123($38.7\%$)	&86($27.0\%$)		\\
	$K\!=\!7$	&216($68.4\%$) &107($33.9\%$)&55($17.4\%$)&221($69.5\%$) &138($43.4\%$)	&96($30.2\%$)		\\
	$K\!=\!8$	&232($73.4\%$) &113($35.8\%$)&64($20.3\%$)&239($75.1\%$) &146($45.9\%$)	&80($25.2\%$)		\\
	$K\!=\!9$	&231($73.1\%$) &112($35.4\%$)&67($21.2\%$)&262($82.4\%$) &142($44.7\%$)	&98($30.8\%$)		\\					
	$K\!=\!10$	&220($69.6\%$) &127($40.2\%$)&66($20.9\%$)&245($77.1\%$) &138($43.4\%$)	&93($29.2\%$)		\\		
	\tableline
	\end{tabular}
	\label{tab:numgcsover}
\end{table*}

~\cite{tal2009} report the presence of two low surface brightness structures, a NW tidal arm and a faint 
SE ``fan''  in NGC4261. These structures are located within the HST observations used by B+12 to extract 
the catalog of GCs, but within the central avoidance region not used in this study (see Section~\ref{sec:data}). 
We have compared the spatial distribution of this faint optical features with the large scale shape
of the GCs distribution in the region of the galaxy closest to the avoidance region, but we have not noticed
any resemblance or clear correlation between the two different structures.

\begin{figure*}[]
	\includegraphics[height=8cm,width=8cm,angle=0]{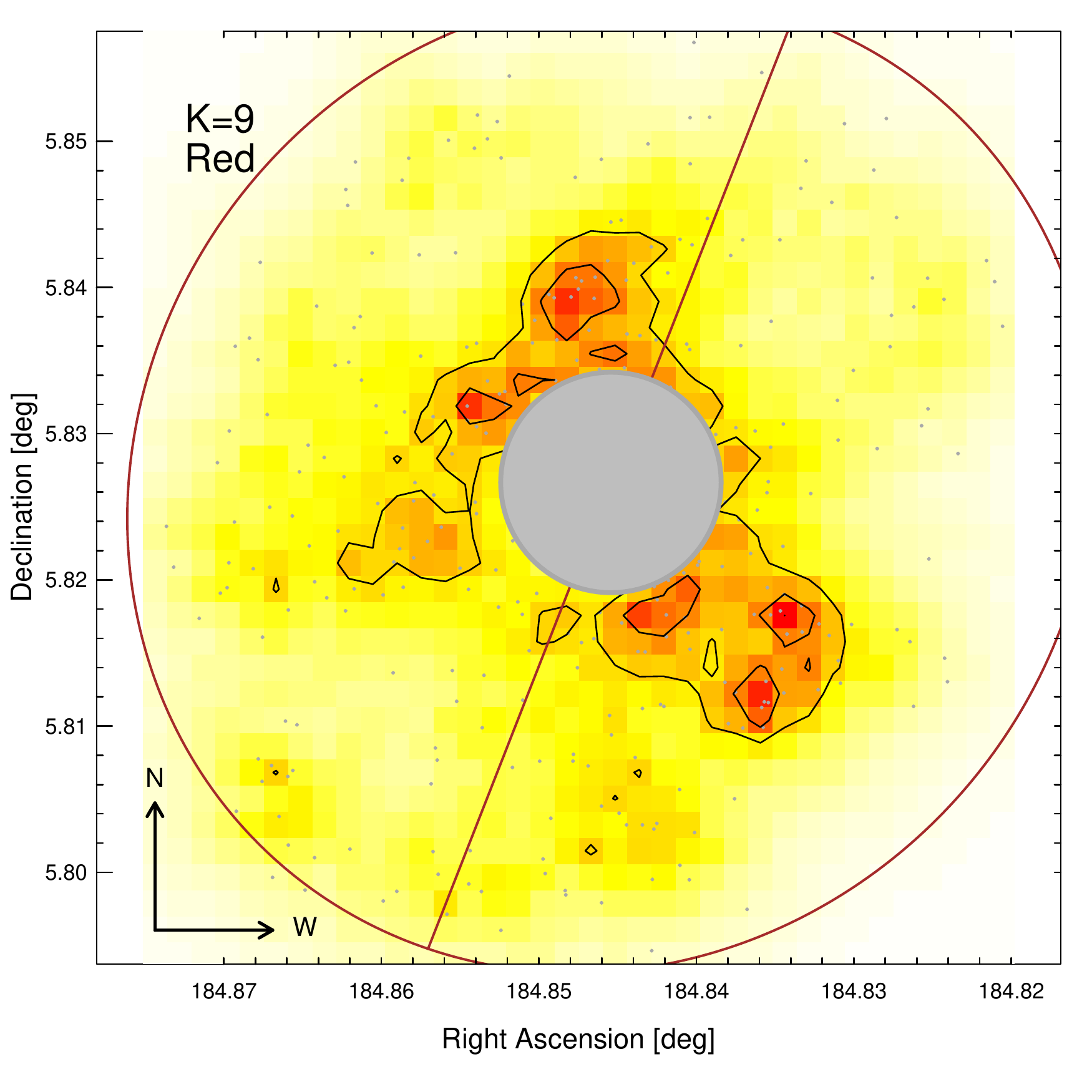}
	\includegraphics[height=8cm,width=8cm,angle=0]{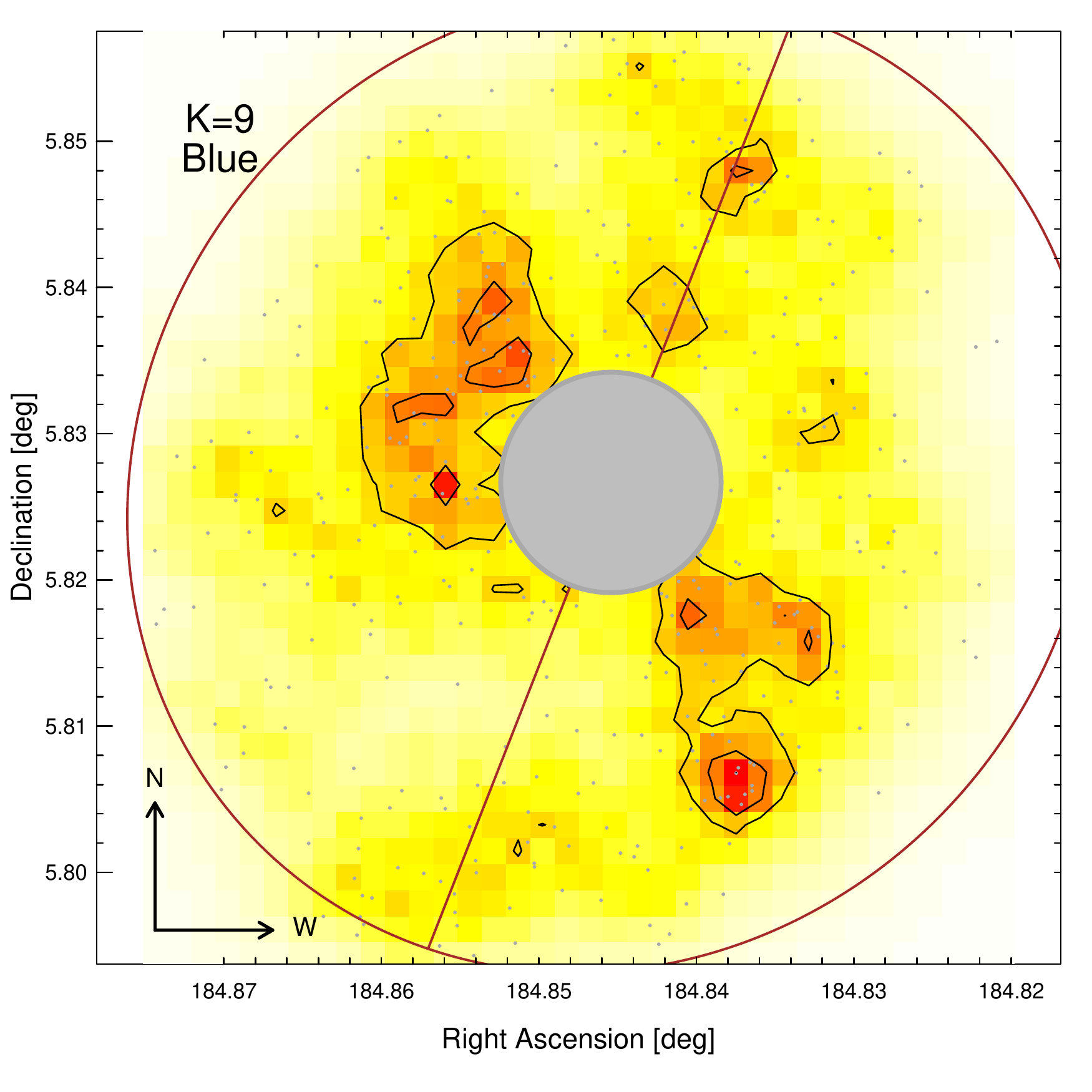}\\	
	\includegraphics[height=8cm,width=8cm,angle=0]{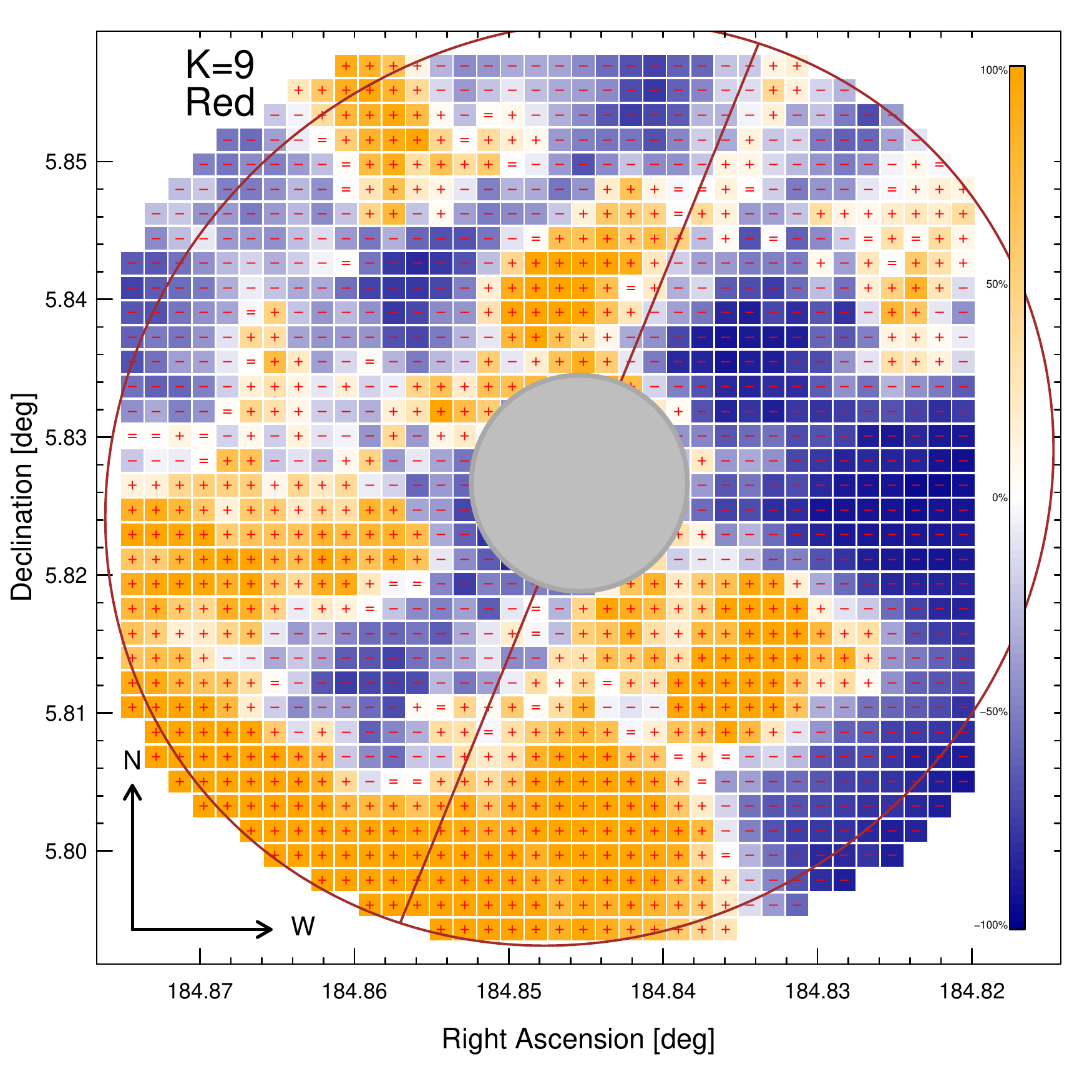}
	\includegraphics[height=8cm,width=8cm,angle=0]{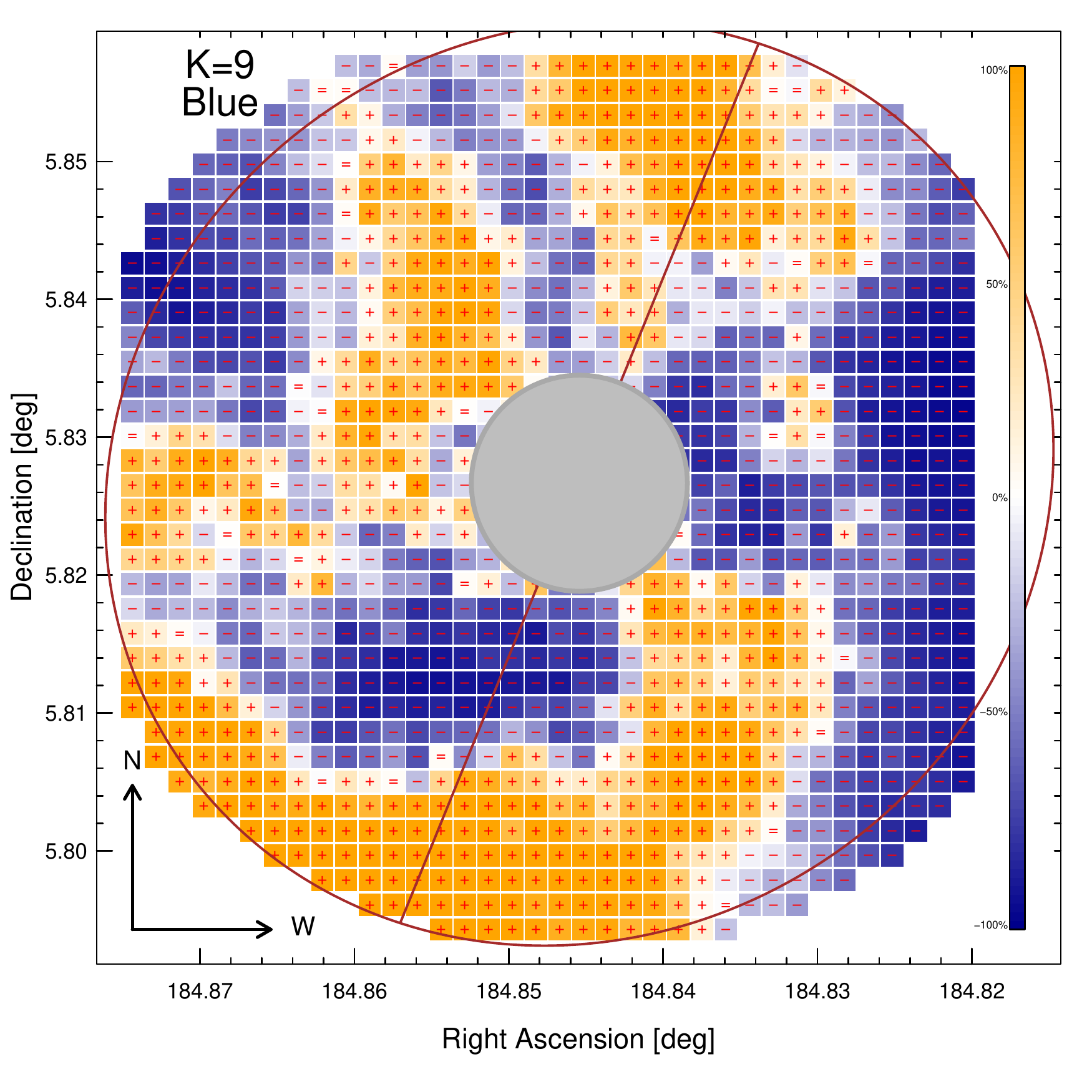}\\	
	\caption{Upper panels: From left to right, observed density maps of red and blue samples for $K\!=\!9$. 
	Lower panels: From left to right, residuals maps of red and blue samples obtained for $K\!=\!9$. The 
	small $+, -$ and $=$ signs within each pixel indicate positive, negative or null residuals respectively.}
	\label{fig:2dmapsredbluengc4261}
\end{figure*}

The $K\!=\!9$ maps of red and blue GC classes (Section~\ref{sec:data}), show significant 
differences in the over-densities, although the overall residuals follow those of the entire sample
(Figure~\ref{fig:2dmapsredbluengc4261}). In particular, a stronger concentration of $\!\geq\!3\sigma$ 
over-dense pixels can be seen in the north of the blue GC residual map. These results do not 
depend on the 
value of the B+12 color threshold used to define blue and red subpopulations, as
long as the threshold is chosen within the interval $V\!-\!I\!=\![1, 1.3]$. The density and residual
maps of the two color classes using ten regularly spaced threshold values in the above interval 
are self consistent. For values outside this interval, the significance of the residual map for one of the 
classes degrades rapidly because of the small number of GCs.

\begin{figure*}[]
	\includegraphics[height=9cm,width=9cm,angle=0]{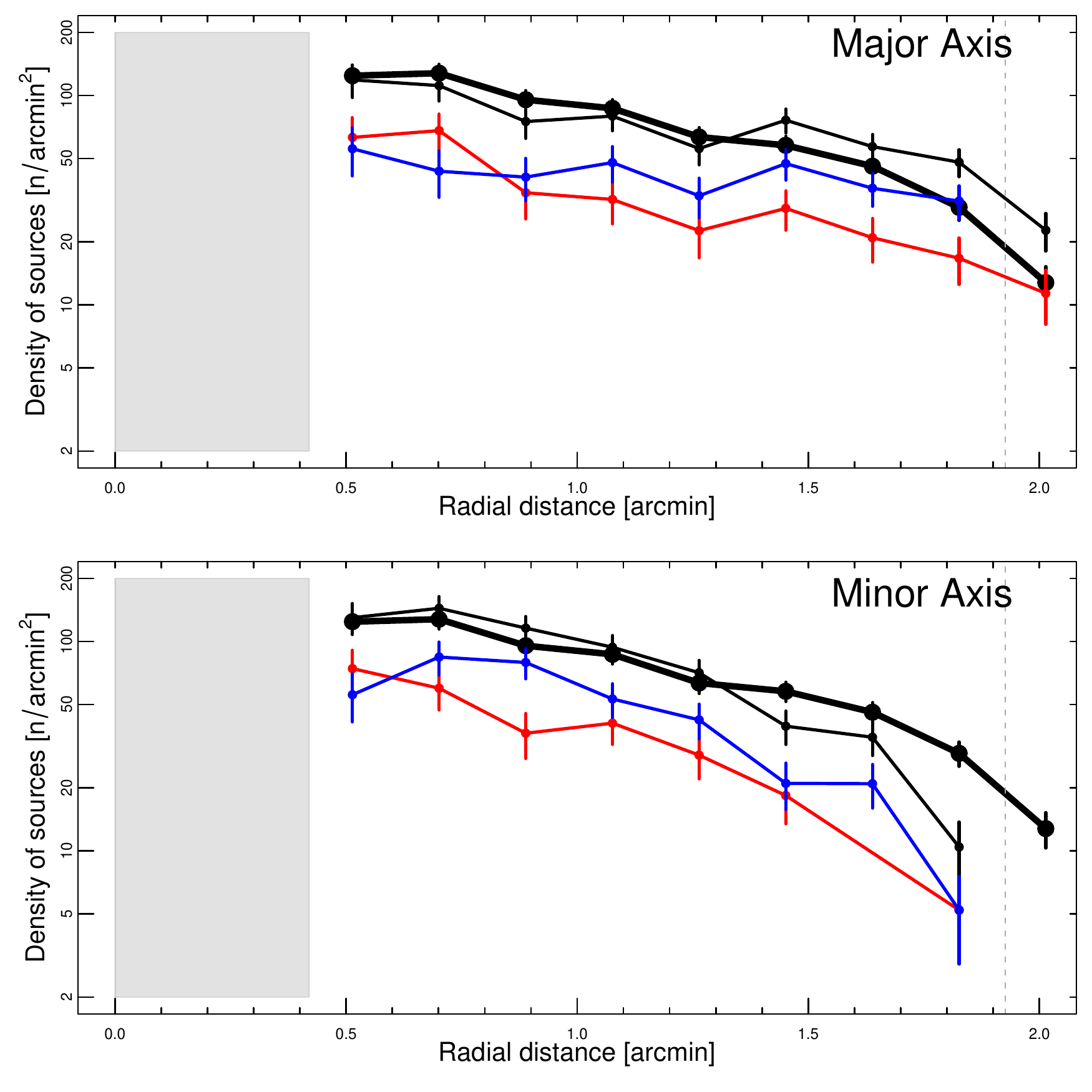}
	\includegraphics[height=9cm,width=9cm,angle=0]{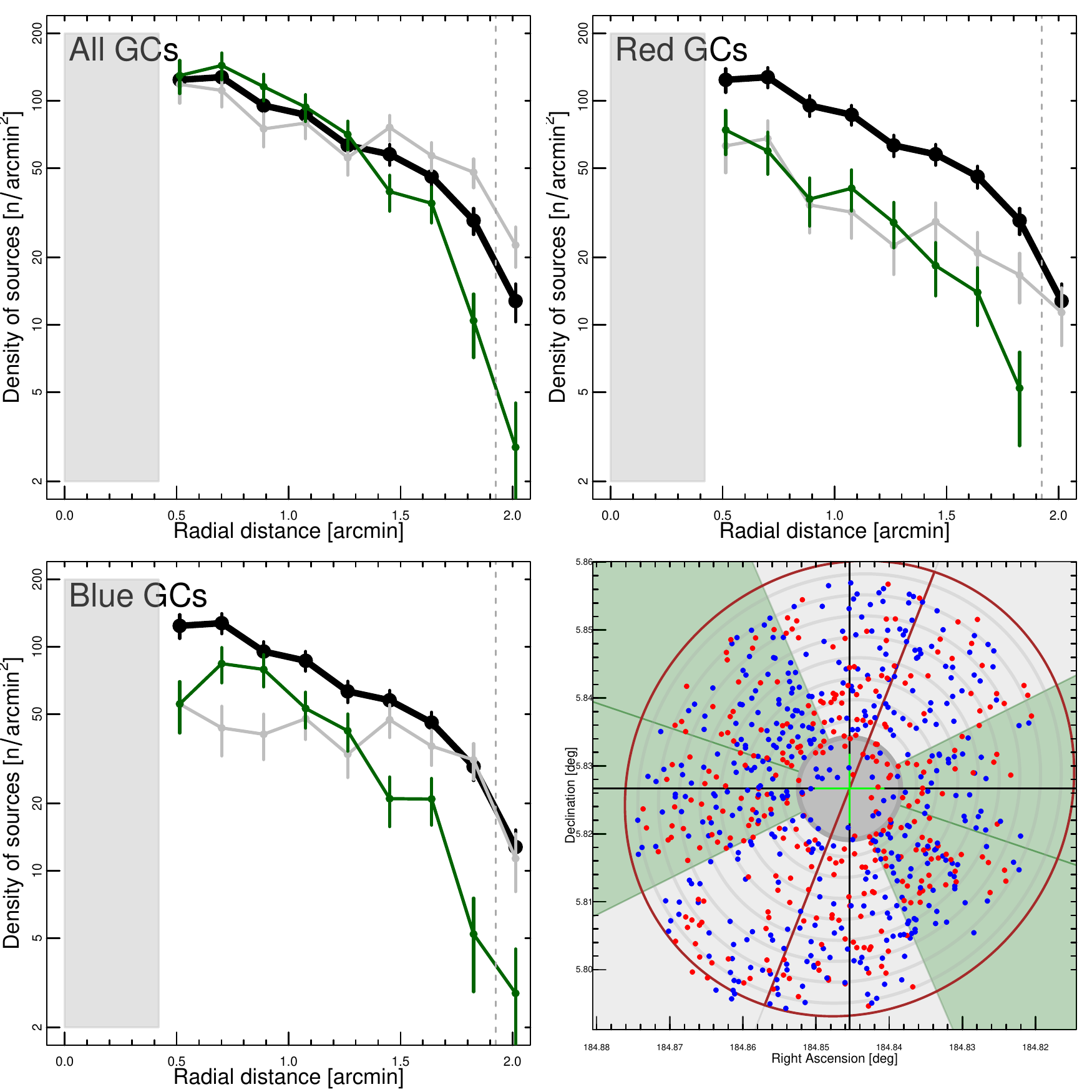}	
	\caption{Left: Radial density profiles of all (thin black line), red (red line) and blue (blue line) GCs calculated
	in elliptical annuli for two azimuthal wedges containing the major and minor axes (upper and 
	lower plot respectively), compared to the overall radial profile of the entire sample (thick black line).
	Middle and right: observed radial density profiles for different types of GCs in the 
	two regions shown in the lower-right panel. The thick black line represents the radial density profile for 
	all GCs integrated over the whole galaxy. The thin gray and green lines represent the major and minor
	axis density profiles respectively.}
       \label{fig:radialdensityprofilesngc4261}
\end{figure*}

The observed radial density profiles of the GC distributions in wedges aligned along the axes of the 
galaxy (Figure~\ref{fig:radialdensityprofilesngc4261}) indicate that GCs have different radial distributions along the 
directions of the major and minor axes of the galaxy. The density profile of the entire sample along the major 
axis is flatter than the average density profile for radii $r\!\geq\!1.2^{'}$, while the minor axis density profile is 
significantly steeper (Figure~\ref{fig:radialdensityprofilesngc4261}, left), after correcting for geometrical effects.
Both red and blue GC profiles are 
more extended along the major axis, compared with the minor axis distributions. This difference is more 
striking for the blue GCs, which have a flat  major axis profile while the minor axis profile first increases at 
small radii, and then plummets towards $D_{25}$.

\begin{figure}[h!]
	\includegraphics[height=9cm,width=9cm,angle=0]{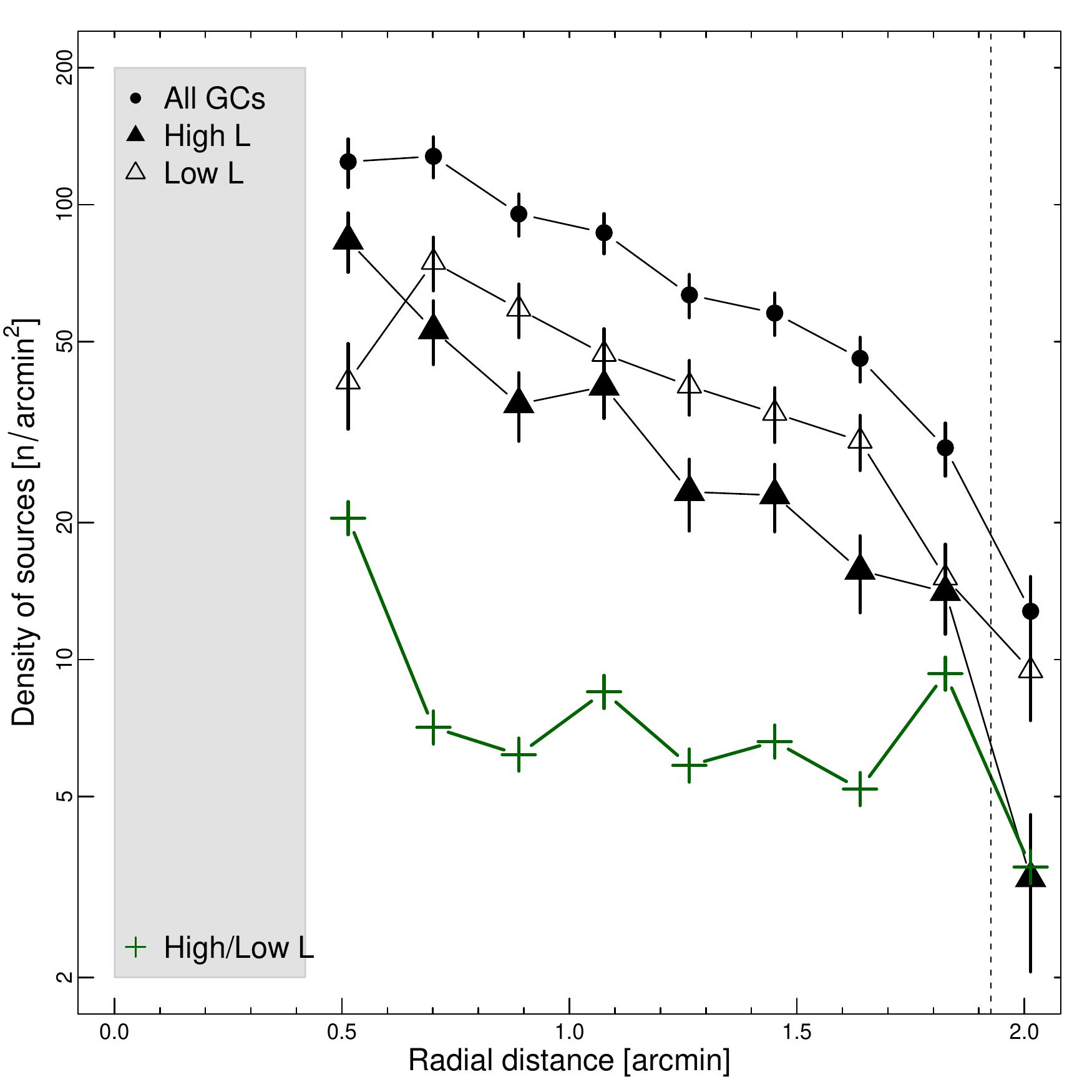}	
	\caption{Observed radial density profiles 
	integrated in elliptical annuli for all (solid circles), high-luminosity (solid triangles) 
	and low-luminosity (open triangles) GCs in NGC4261. The green symbols represent the ratio of the high-luminosity
	to low-luminosity densities profiles, multiplied by 10 for display purposes. In both plots the gray 
	area corresponds to the 
	center of NGC4261 excluded from our analysis where the GC detections are highly 
	incomplete (see (B+12)). The vertical dotted lines represent
	the major axis of the $D_{25}$ isophote.}
       \label{fig:radialhighllowl}
\end{figure}

\begin{figure*}[!h]
	\includegraphics[height=9cm,width=9cm,angle=0]{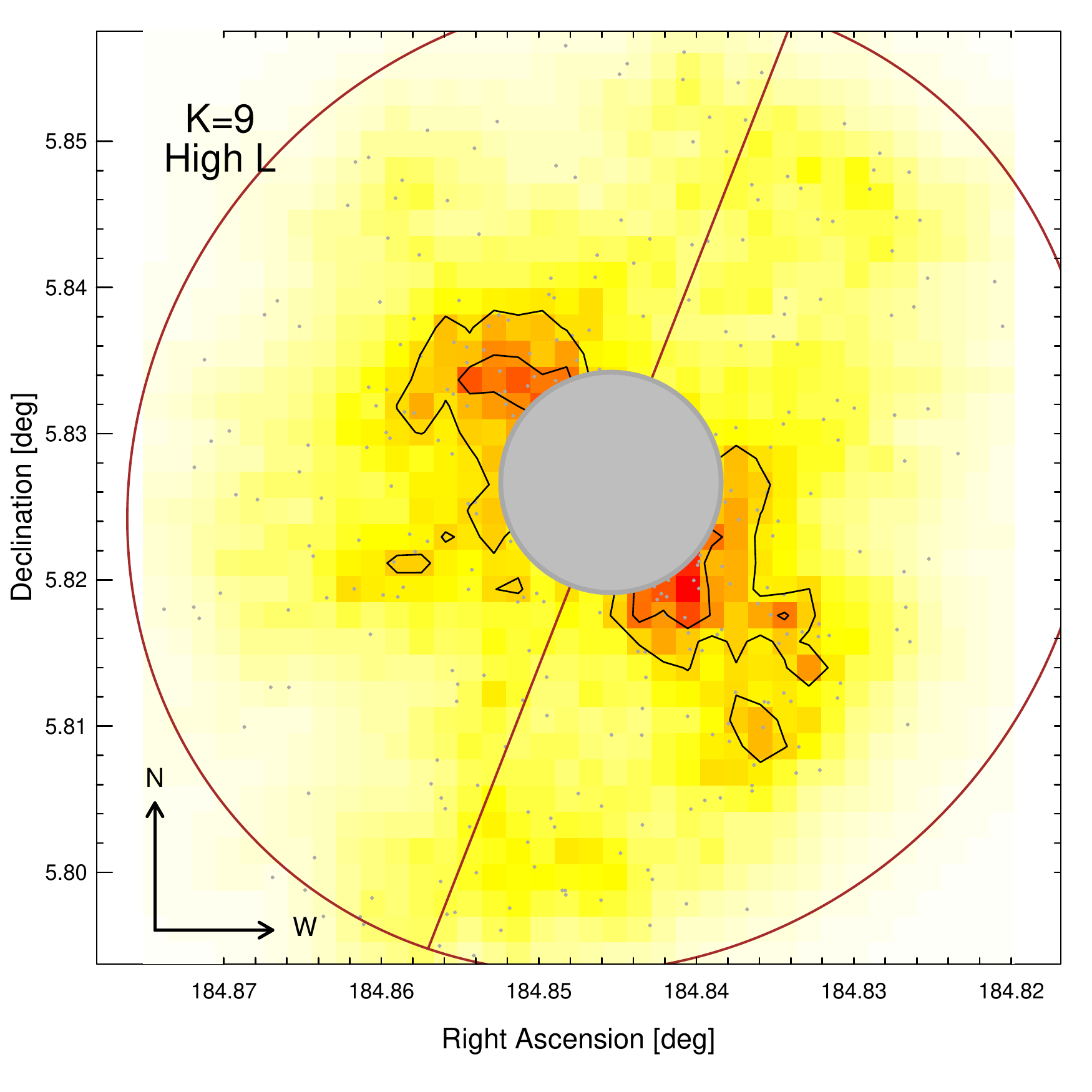}
	\includegraphics[height=9cm,width=9cm,angle=0]{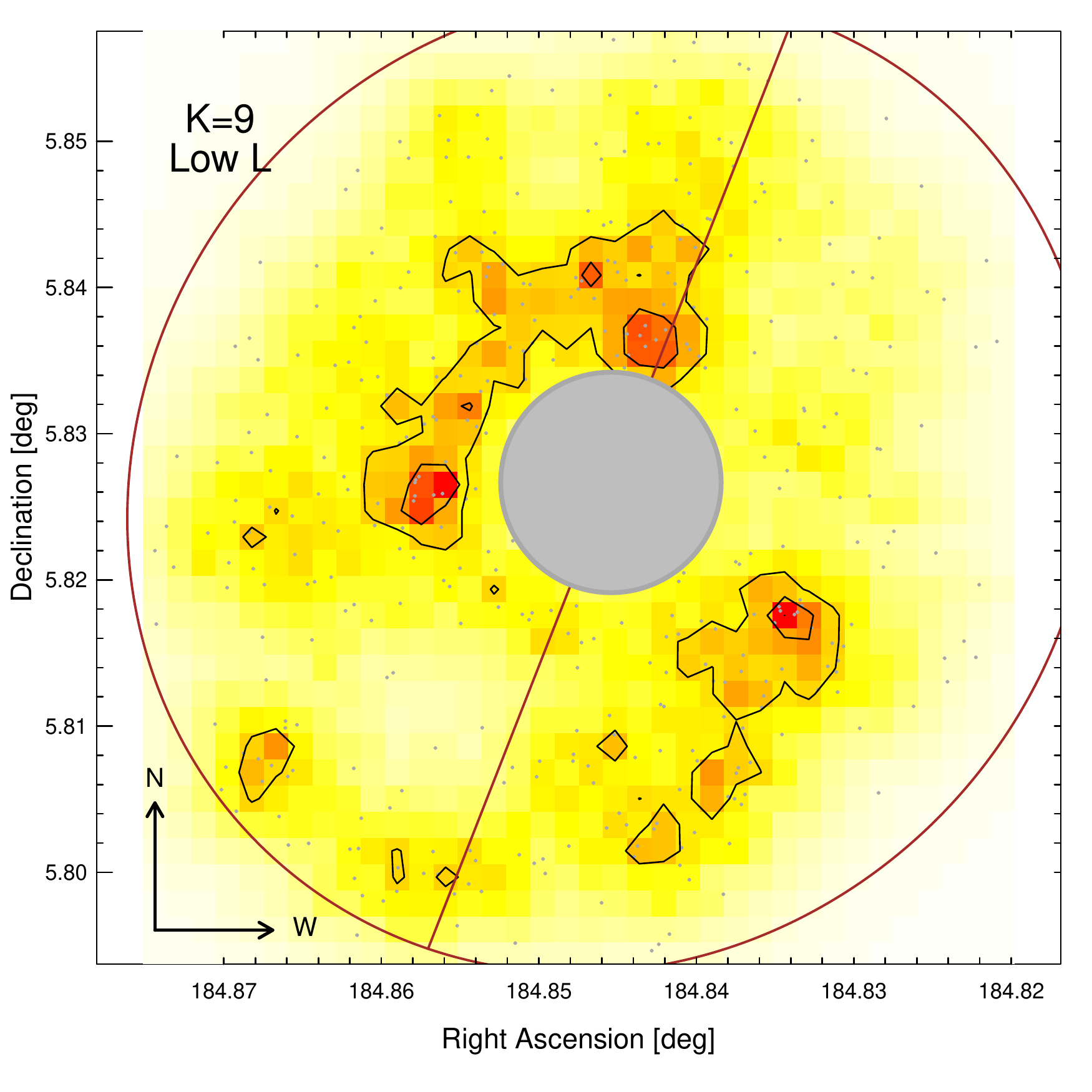}\\
	\includegraphics[height=9cm,width=9cm,angle=0]{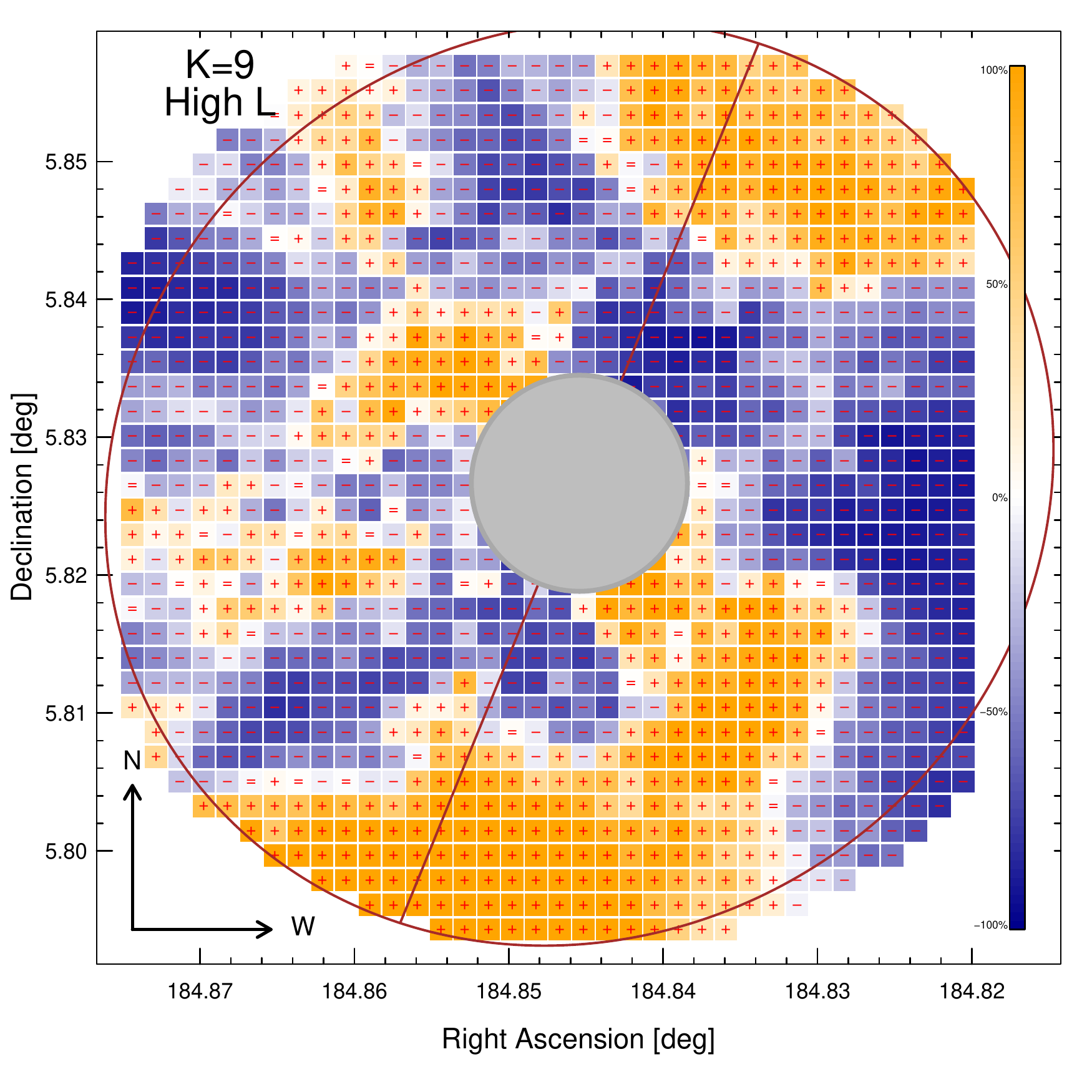}
	\includegraphics[height=9cm,width=9cm,angle=0]{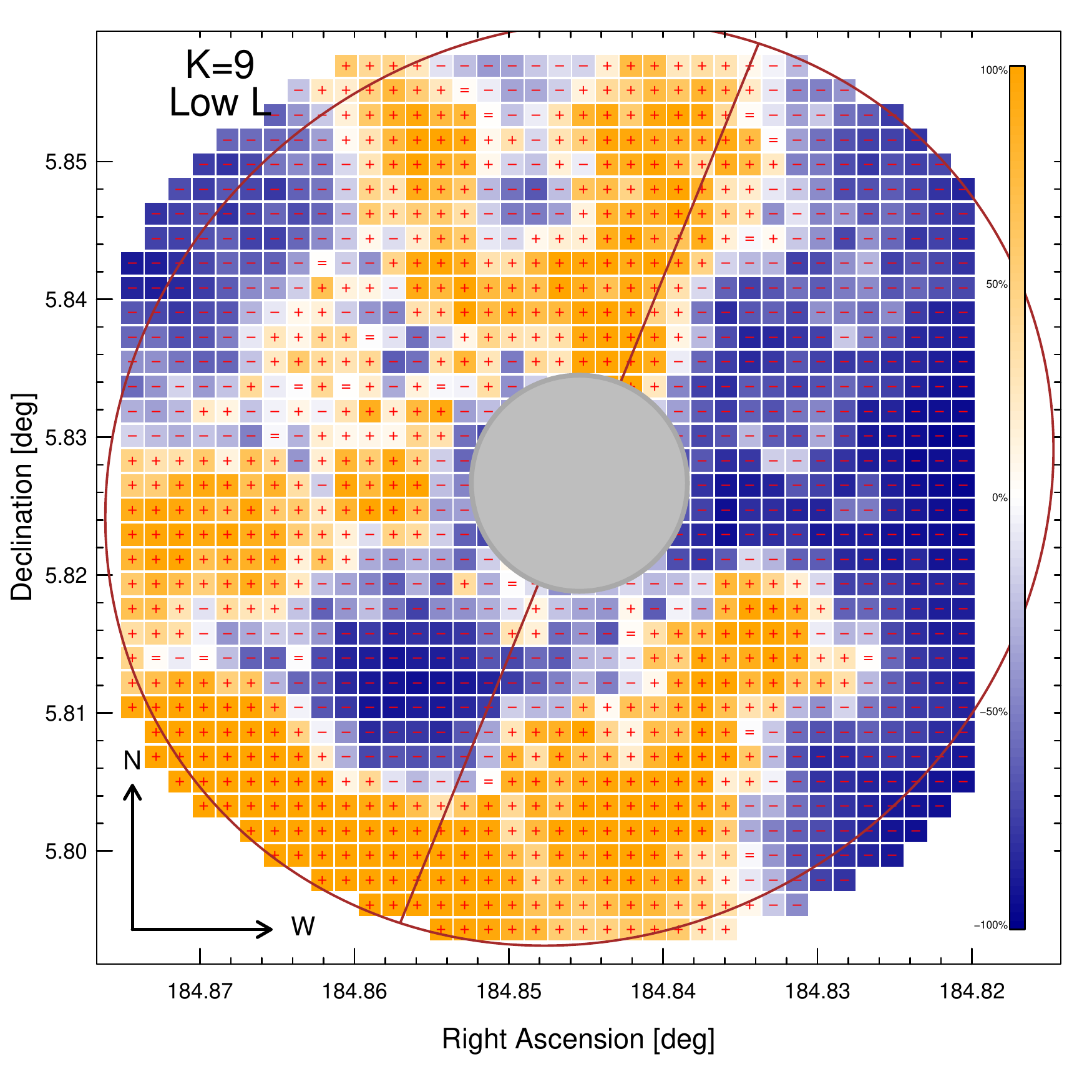}\\		
	\caption{Upper panels: From left to right, $K\!=\!9$ density maps derived from the spatial distribution 
	of 316 high-luminosity GCs ($I\!<\!23$) and 402 low-luminosity GCs with $I\!\geq\!23$. Lower panels: 
	From left to right, $K\!=\!9$ residual plots of high and low luminosity samples. In both plots the gray area 
	corresponding to the center of NGC4261 excluded from our analysis and the $D_{25}$ elliptical isophote of the 
	galaxy are shown for reference. The small $+, -$ and $=$ signs within each 
	pixel indicate positive, negative or null residuals respectively.}
	\label{fig:2dmapshighlowngc4261}
\end{figure*}

\begin{figure*}[]
	\includegraphics[height=9cm,width=9cm,angle=0]{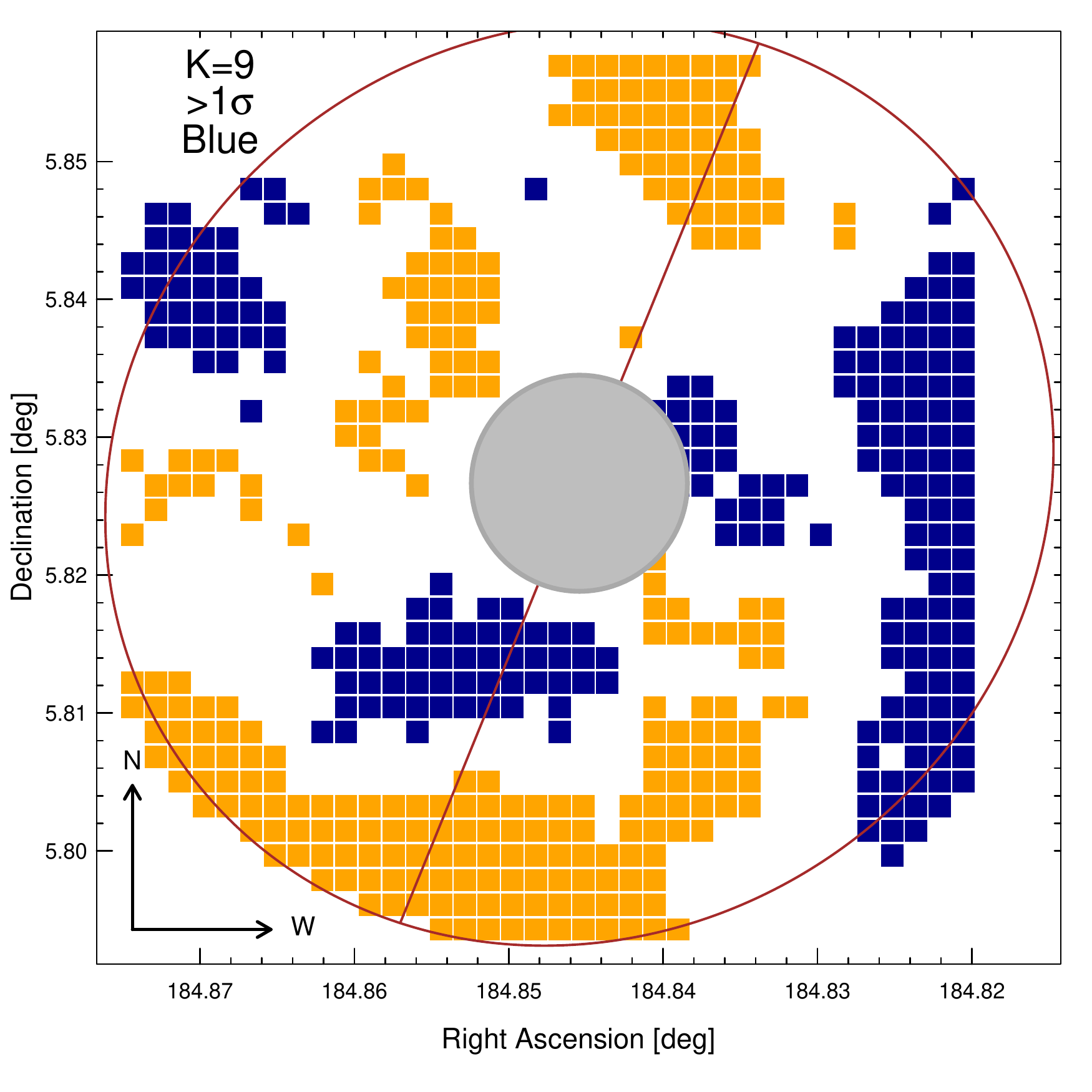}
	\includegraphics[height=9cm,width=9cm,angle=0]{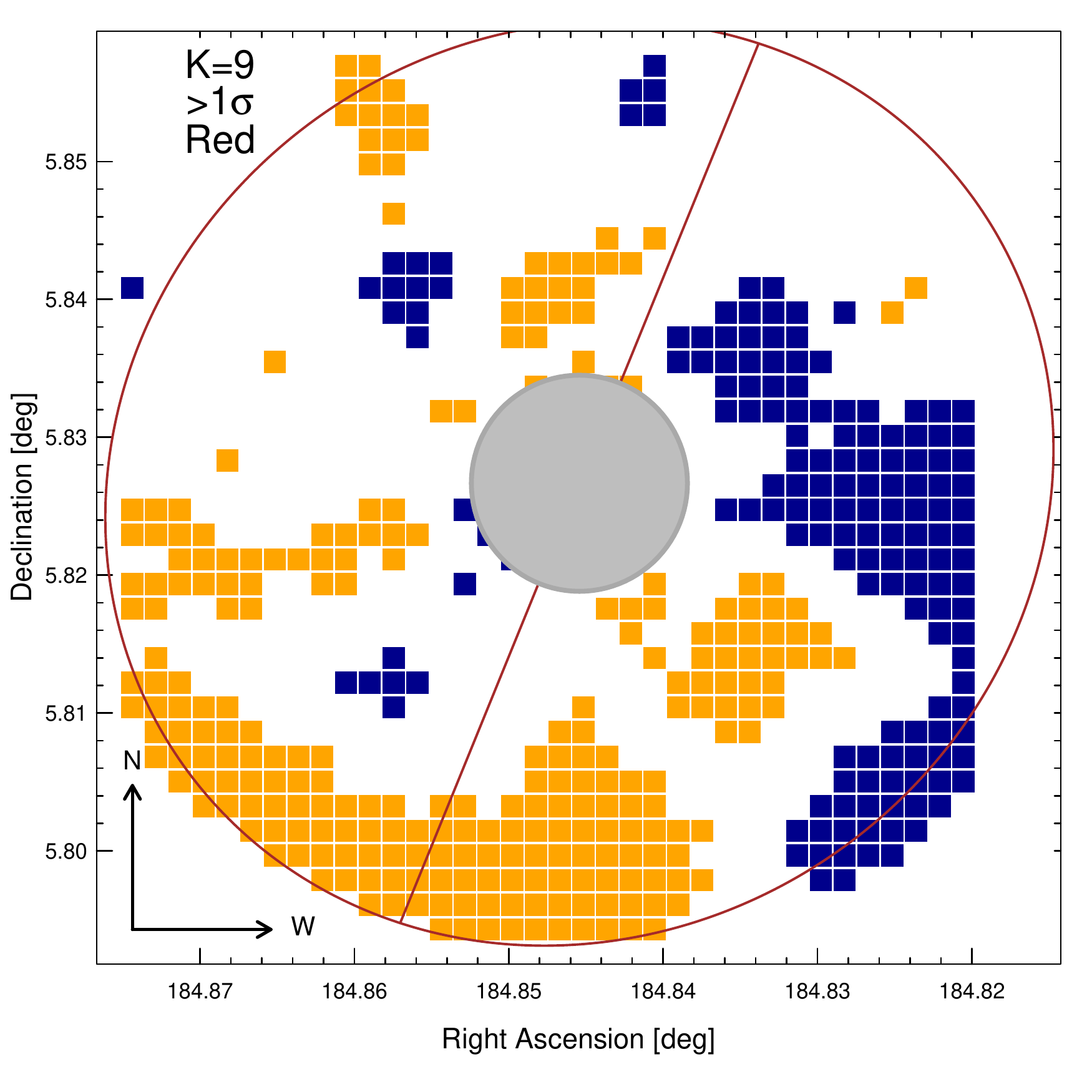}\\
	\includegraphics[height=9cm,width=9cm,angle=0]{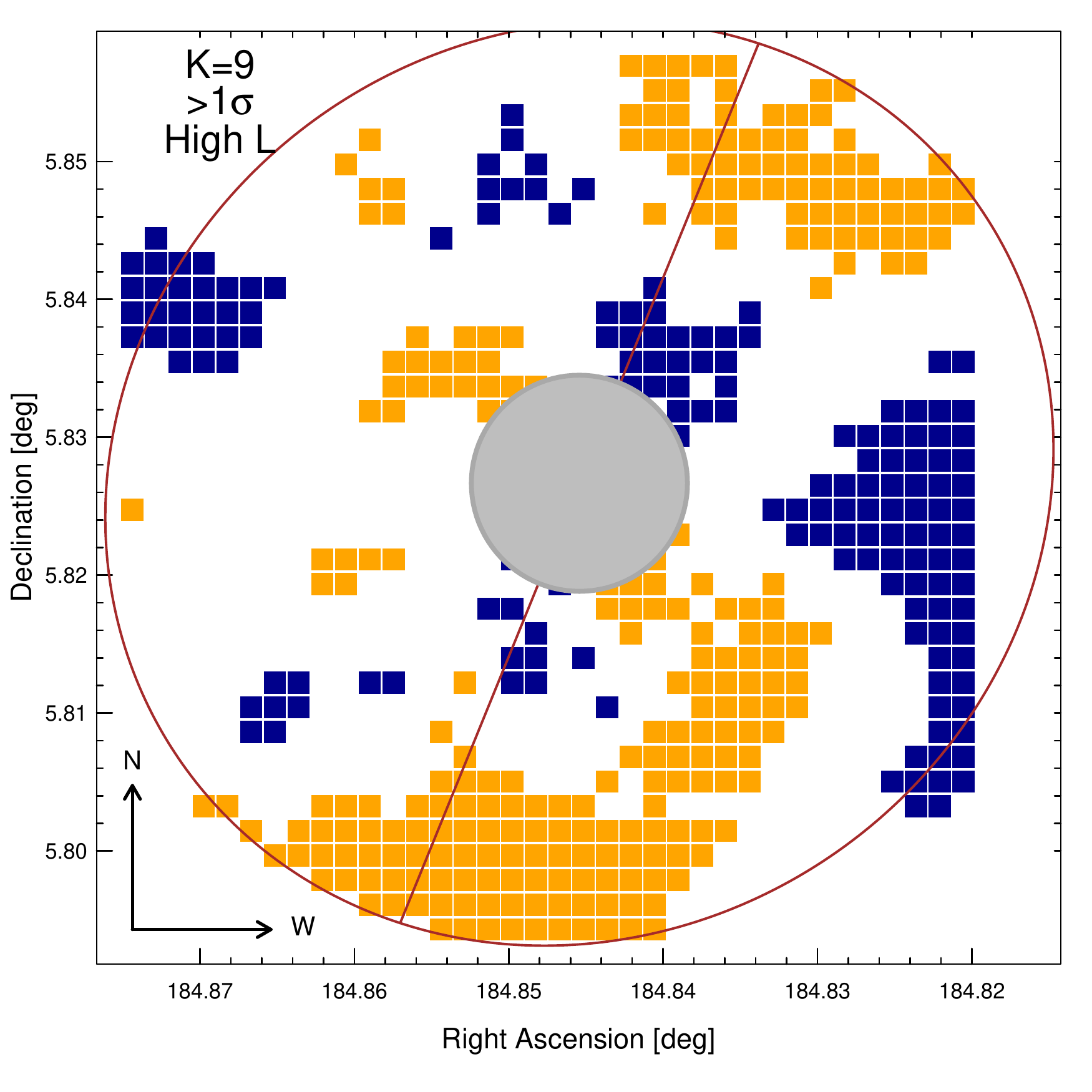}
	\includegraphics[height=9cm,width=9cm,angle=0]{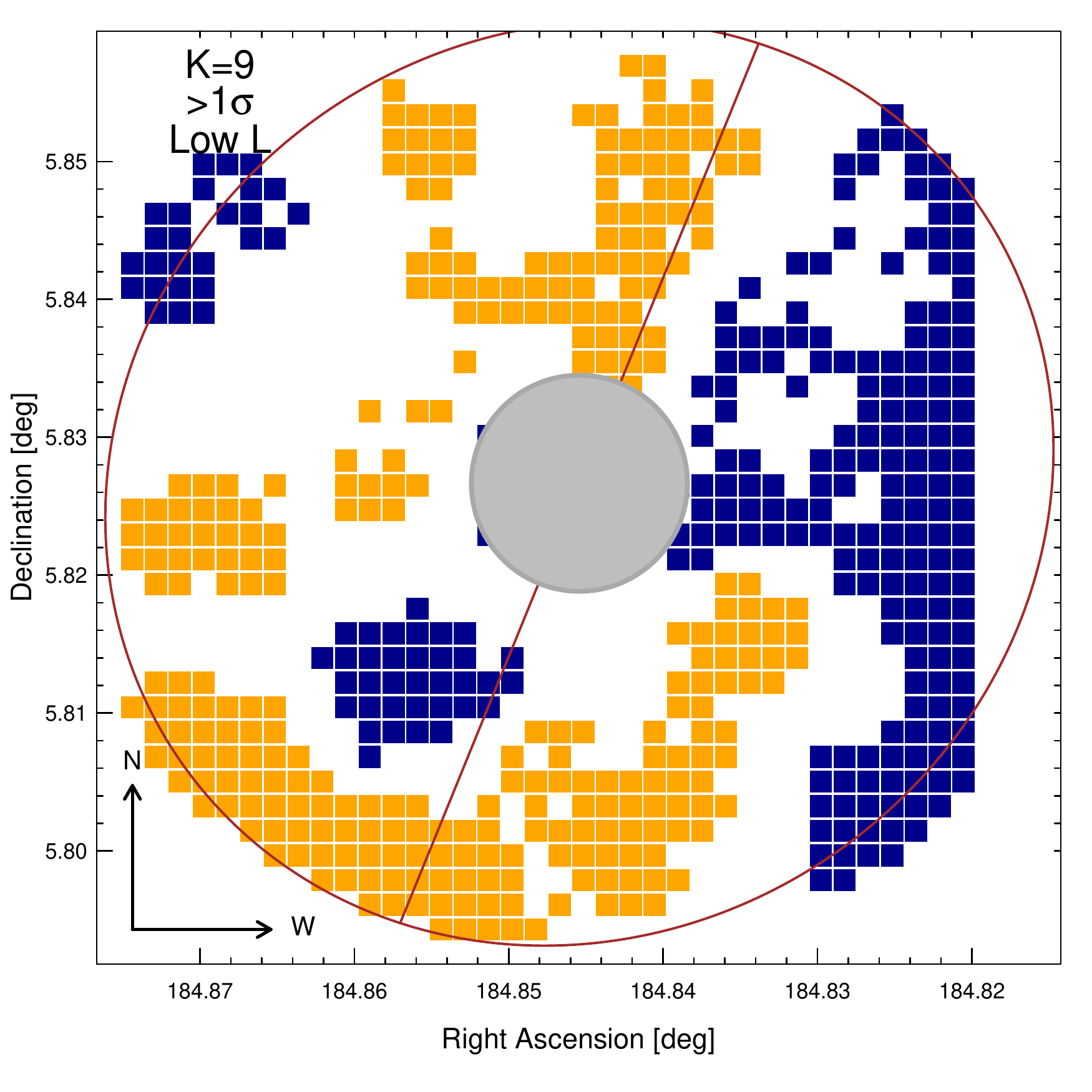}\\
	\caption{Upper Panels: positions of the $\!>\!1\sigma$ residuals for blue (left) and red (right) samples of 
	GCs. Lower panels: positions of the $\!>\!1\sigma$ residuals for high-luminosity (left) and low-luminosity 
	(right) samples of GCs. In all plots the gray area corresponding to the center of NGC4261 excluded from 
	our analysis and the $D_{25}$ elliptical isophote of the galaxy are shown for reference.}
	\label{fig:2dmapshighlowngc4261sigmas}
\end{figure*}

The $K\!=\!9$ density maps for the high and low luminosity subsamples (Table~\ref{tab:summary}) 
are shown in Figure~\ref{fig:2dmapshighlowngc4261}. The density distributions show that the 
high-L GCs are more centrally concentrated than the low-L ones. This is also reflected by the azimuthally 
averaged radial profiles (Figure~\ref{fig:radialhighllowl}). This effect could be at least in part 
due to incompleteness that will be more pronounced in the inner radii where the galaxy stellar light is more 
intense. Although there is a small radial dependence, this effect is minimized by only using GCs with 
$I\!\leq\!24$ for which, according to B+12, a completeness of 75\% is achieved. However, 
there are azimuthal differences at the same radius that cannot be due 
to incompleteness effects. Figure~\ref{fig:2dmapshighlowngc4261sigmas} (lower panels) shows 
the position of the pixels with residuals larger than $1\sigma$ obtained using $K\!=\!9$ for high-L 
and low-L GCs (the same plots for blue and red GCs are shown in the upper panels of
Figure~\ref{fig:2dmapshighlowngc4261sigmas} for reference). 
These plots also show significant differences: there is an excess of High-L GCs to the W of the major 
axis in the N, while the Low-L GCs present an excess to the opposite SE side. The same 
conclusions about the spatial distributions of low and high luminosity classes of GCs 
can be drawn using different values of the threshold magnitude within the interval 
$I\!=\![22, 23.5]$, using ten regularly spaced magnitude thresholds. 
The High-L GCs are, in each case, more centrally concentrated than 
Low-L GCs, and High-L residuals show a unique significant positive density enhancement in the N-W 
region of the galaxy. Other values of the $I$ magnitude 
have not been considered because they would generate luminosity classes too unbalanced to 
correctly estimate the significance of the results.

\subsection{Color vs Spatial distribution of GCs}
\label{subsec:color}

\begin{figure}[h]
	\includegraphics[height=8.5cm,width=8.5cm,angle=0]{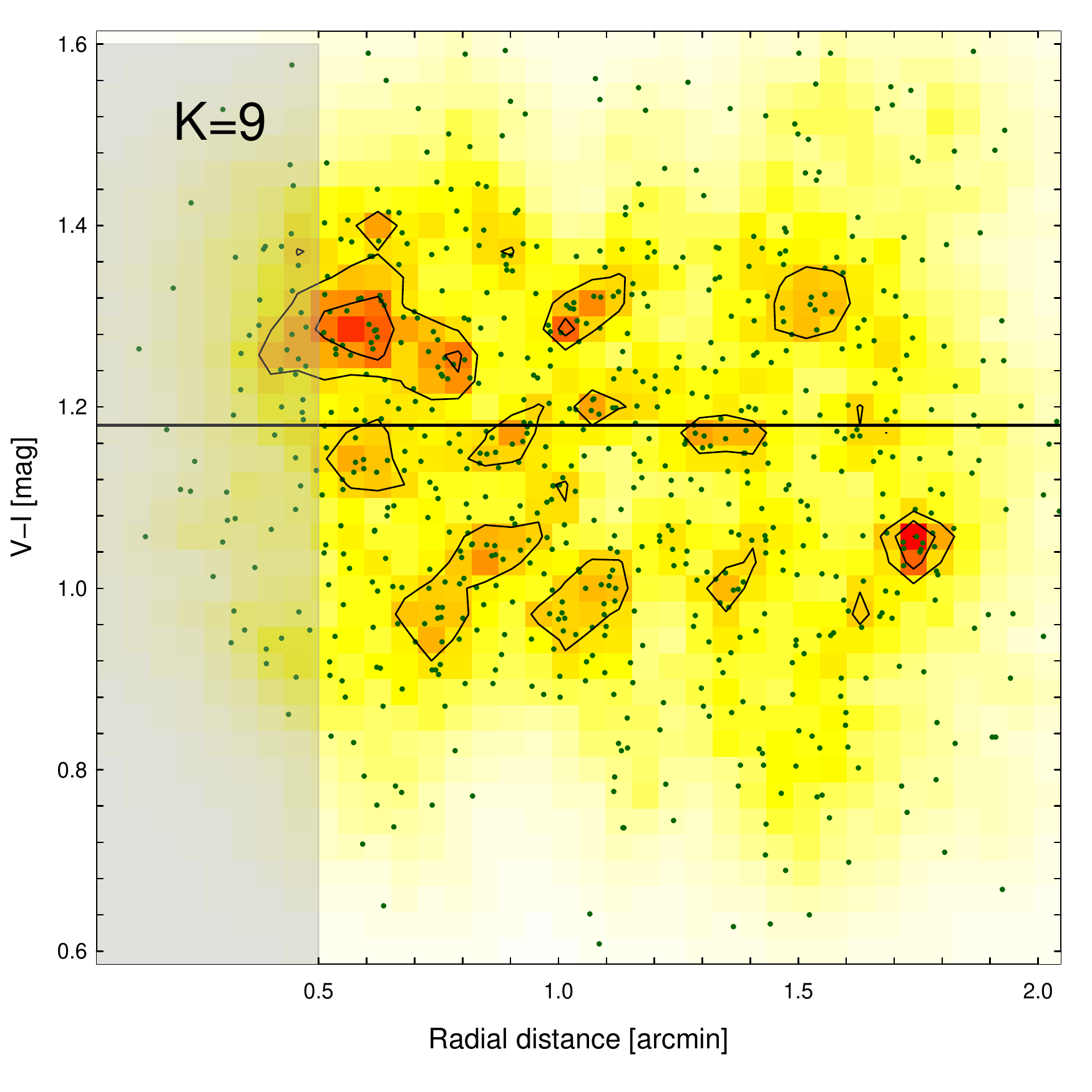}
	\caption{Density map of GCs distribution in the plane generated by the radial distance
	and the $V\!-\!I$ color obtained with the KNN method for $K\!=\!9$. The horizontal black line shows 
	the color value used as threshold to separate Red and Blue GCs in this paper. The gray area corresponds 
	to the center of NGC4261 excluded from our analysis. The green points represent the observed positions 
	of the GCs used to determine the density.}
	\label{fig:mapcolorclasses}
\end{figure}

\begin{figure}[h]
	\includegraphics[height=8.5cm,width=8.5cm,angle=0]{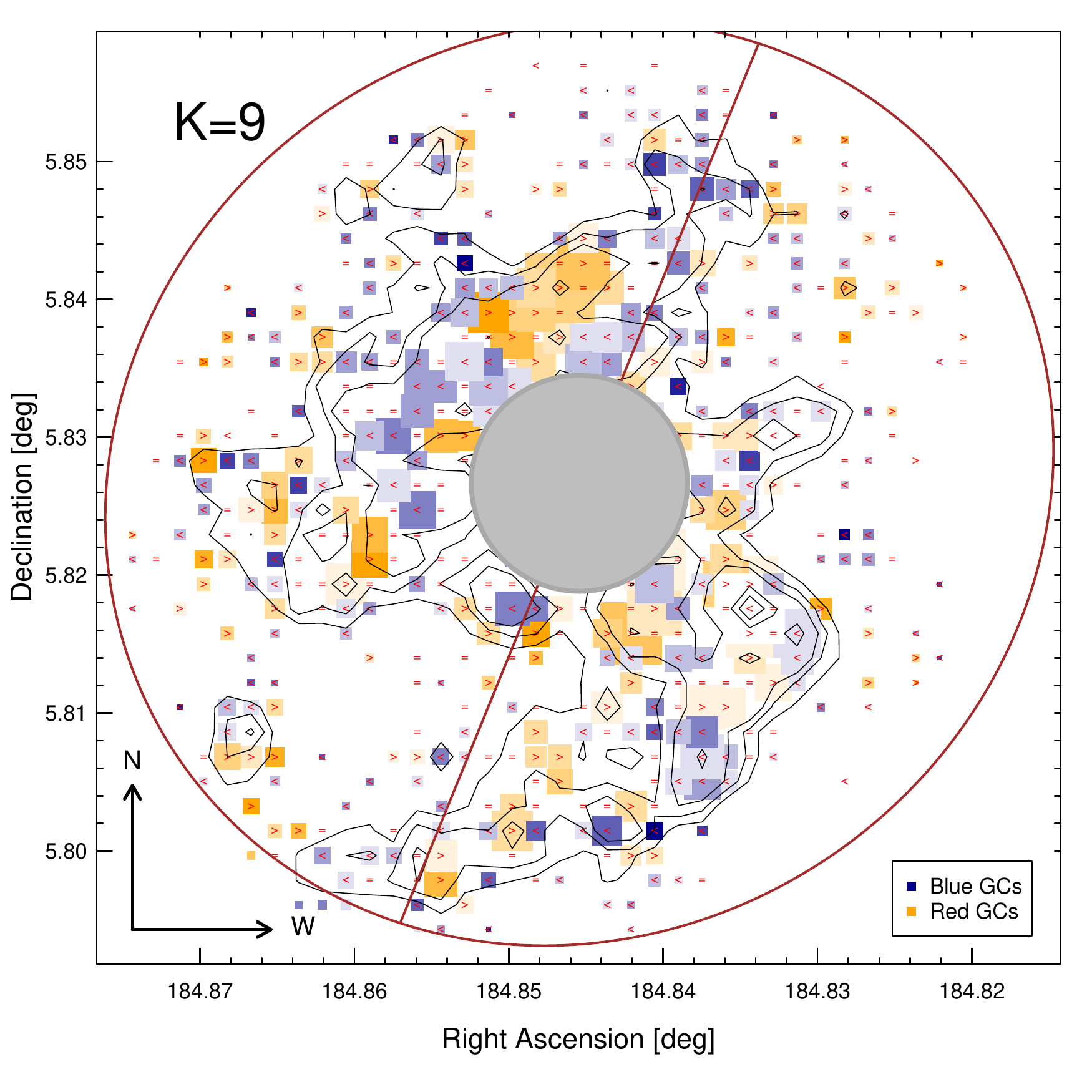}	
	\caption{Map of the average color of GCs in pixels. The size of the symbols is proportional to 
	the density of GCs distribution
	evaluated with the KNN and $K\!=\!9$ and the color intensity of the symbols is proportional to the 
	difference between the color threshold value $V\!-\!I\!=\!1.18$ and the average color of the GCs in each
	pixel. Small red symbols $>$, $=$ and $<$ are drawn within each pixel with average
	GCs color $V\!-\!I\!>\!1.18$, $V\!-\!I\!=\!1.18$ and $V\!-\!I\!<\!1.18$ respectively. The isodensity 
	contours reflect the density map with $K\!=\!9$ derived from the distribution of all GCs.}
	\label{fig:mapcolorclasses2}
\end{figure}

 \cite{blom2012} found significant effects in the color vs galactocentric distance distribution 
of the GC system of the galaxy NGC4365. We have investigated if similar effects are present in 
NGC4261. We have produced the density maps of the distribution of GCs in the 
radial distance vs $V\!-\!I$ color with the KNN method for $K\!=\!\{5, 6, 7, 8, 9, 10\}$. The map for $K\!=\!9$ 
is shown in Figure~\ref{fig:mapcolorclasses}. The clumpy distribution of over-densities does 
not show clear separation between different classes of color except for a tendency of red GCs to be 
more centrally concentrated than blue GCs, in agreement with the density and residual maps of the 
two separate 
color classes of GCs (Figure~\ref{fig:2dmapsredbluengc4261}). In particular, at large radial distances, 
the only interesting density enhancement of Figure~\ref{fig:mapcolorclasses} is located in  
$r\!\sim\!1.75'$ and $V\!-\!I\!\sim\!1.12$. This feature corresponds to the sum of the two significant density 
enhancements in the spatial density map of the blue GCs class located along the major axis of the 
galaxy in the N-W quadrant and in the S-W quadrant (see upper left plot in 
Figure~\ref{fig:2dmapsredbluengc4261}). A more detailed representation of the 
GCs color distribution in NGC4261 as a function of the position in the galaxy can be seen in 
Figure~\ref{fig:mapcolorclasses2}, where each pixel is colored according to the average color of the 
GCs placed within the pixel. The size of the pixels is proportional to the density evaluated with the KNN
method for $K\!=\!9$ and the density contours are derived from the same density map. 
Figure~\ref{fig:mapcolorclasses2} confirms that red and blue GCs tends to follow the same overall 
spatial patterns, except for the different small differences visible in the upper panels in 
Figure~\ref{fig:2dmapsredbluengc4261}, obtained using the $V\!-\!I\!=\!1.18$ color threshold value. 

\section{Discussion}
\label{sec:discussion}

Our study of the 2D projected distribution of GCs in the elliptical galaxy 
NGC4261 by means of KNN density maps has confirmed at high statistical significance the 
presence of the 2D large-scale anisotropy reported 
by B+12. The morphology of this anisotropy suggests a large-scale 
spiral-like over-density which originates along the major axis and follows the $D_{25}$ elliptical isophote 
in the periphery of the galaxy. Alternatively, there may be a system of 
broken shells enclosing the major axis in both directions roughly along the $D_{25}$ ellipse, with 
local density enhancements especially along the northern segment of the galaxy major axis. 
We find the same overall ``grand pattern'' when the sample is 
subdivided by either color or luminosity, although there are specific differences between the subsamples. 
In particular, while the southern feature is seen in both red and blue GCs, the northern anisotropy 
shows some segregation of red and blue GCs. Similarly the morphology of the residuals differs in the 
N-W (see Section~\ref{subsec:density}) suggesting a lack of high-L GCs in the inner major axis. 
Overall there are more red GCs at smaller radial distances (see Figure~\ref{fig:radialdensity}). 
However, we note that these effects may be correlated, since the high-L sample has a slightly larger 
fraction ($\sim\!56\%$) of blue GCs (see Table~\ref{tab:summary}).

At small radial distances, the density enhancements of the GCs distribution are
aligned along the minor axis of the galaxy. The overall distribution is driven by high-L GCs which
are located along the minor axis direction while low-L GCs do follow the broken-shell/spiral arm 
features more closely. These results differ significantly from recent results from~\cite{wang2013}, 
which show that GCs of the early-type galaxies in the ACS Virgo Cluster survey tend to be
aligned along the major axis in galaxies with visible elongation and intermediate to high luminosity. 
The GCs system in NGC4261 shows that the geometry of the GC distribution observed in 
the~\cite{wang2013} sample is not universal, even though the different conclusions based on the 
results provided by our method may depend on the rich sample of NGC4261 GCs used which, 
nonetheless, does not cover the whole GCs system.

As already discussed in B+12, these large-scale features suggest global shaping events, 
however their nature is not clearly understood. Tidal tails and shell-like ripples of the stellar surface 
brightness have been reported in several elliptical galaxies, and have been modeled in terms of mergers 
and interactions with satellites 
(e.g.,~\citealt{schweizer1980};~\citealt{quinn1984};~\citealt{hernquist1988},~\citealt{hernquist1989}). 
However, these simulations did not address GC systems.

In NGC4261, no significant sign of perturbation 
of the stellar surface brightness of the galaxy (except for two faint central structures noticed 
by~\citealt{tal2009}), is observed, indicating 
that a major merging event has not likely occurred in the last $\sim$1 Gy. 
{However,~\cite{ferrarese1996} observed boxy isophotes which are consistent with past merger.
As discussed in B+12, a major merger scenario may be an alternative for the global 2D 
anisotropy of the GC distribution only if the relaxation timescale for the GCs is larger than that of the stars. 
Another possibility in the major merger scenario is that the GC over-densities may be related with GC formed during 
the merger in shocked 
regions and tidal tails, such as the the evolved remnant of the young massive cluster populations detected 
in the Antennae~\citep{whitmore2010} and NGC 7252~\citep{bastian2013}. 

The peculiar spatial distribution of GCs in NGC4261 could also partially be influenced 
the GC systems of the smaller galaxies visible in the NGC4261 field and observed 
by~\cite{tal2009}. In particular, two dwarf galaxies are visible in the HST images 
of NGC4261 GCs. The first, with $B\!=\!16.2$ mag~\citep{binggeli1985},  
is located in the N-W corner of the HST mosaic $(\mathrm{R.A.}\!=\!12\!:\!19\!:\!17.6,
\mathrm{Dec.}\!=\!+5\!:\!52\!:\!39)$ and is too far from NGC4261 to contribute to the NGC4261 GCs 
system. The second ($B\!=\!15.3$) is close to the S-W spiral-arm feature visible in the 
density map of the NGC4261 GCs system $(\mathrm{R.A.}\!=\!12\!:\!19\!:\!23,\mathrm{Dec.}\!=\!+5\!:\!47\!:\!51)$.
Similar galaxies have compact GCs systems (see~\citep{dirsch2005}). Thus, if this galaxy has an 
GCs system overlapping with the NGC4261 GCs population, this would only influence a small 
region of the spiral-like over-density associated to the NGC4261 GCs spatial distribution. 

A third, brighter, galaxy $(B\!=\!13.8)$ $(\mathrm{R.A.}\!=\!12\!:\!19\!:\!35,
\mathrm{Dec.}\!=\!+05\!:\!50\!:\!47)$, could in principle have a GCs population spatially overlapping the 
NGC4261 GCs system. Nonetheless, we notice that this galaxy is located on the E side of 
NGC4261 and no significant density enhancements is visible in the E outskirt of NGC4261
(Figure~\ref{fig:2dmapsngc4261}). A feature resembling a broken shell is apparent in the E 
quadrants of the NGC4261 closer to the center of the galaxy. Assuming a power-law radial 
density profile for the GCs system of the companion galaxy, the result of its contribution to 
the NGC4261 GCs density map would be an enhancement of the density decreasing from the
outskirt to the center of NGC4261. For this reason, the contamination from the GCs system of the
neighboring galaxy could only reduce the significance of the observed over-density in the E
side of NGC4261, and would not change qualitatively the results of our analysis.

We cannot exclude that present and past gravitational interactions with these neighboring galaxies 
may have affected the GC system of NGC4261, but the molding of these effects is beyond the scope 
of this paper.

Recent SPH simulations of the formation of GCs in merging and interactive galaxies
suggest that newly formed metal-rich GCs tend to concentrate at the center of the merger
remnant elliptical galaxy while metal-poor GCs are distributed in the outer parts due to strong angular 
momentum transfer (see~\citealt{bekki2002}). However, our residual maps, while showing 
a few differences in the over-densities of red and blue GCs, demonstrate that the overall 
patterns are similar, as already noted by B+12. This result argues against a ``rejuvenation'' 
of the GC populations that would result in an increase of the higher metallicity red population. 
B+12 also noted that a minor merger scenario cannot 
be excluded, but the incoming galaxy should be particularly GC-rich, and that a displacement 
of the GC population by a galaxy fly-by may be a possible explanation. 
 
The characterization of the distribution of GCs in the radial distance vs $V\!-\!I$ space 
(Section~\ref{subsec:color}) demonstrates the lack of a clear bimodality in the color distribution 
of GCs over the whole interval of radial distances considered in our analysis. Nonetheless, the expected
differences of the spatial distributions of red and blue GCs are evident in our analysis, regardless of 
the specific color value used to separate the two classes (Section~\ref{subsec:density}).

While our paper cannot solve the nature of the perturbations in the spatial distribution of GCs in NGC4261, 
we have contributed detailed 2D maps of the morphology of the 
asymmetry at different spatial scales. The morphology of the perturbations suggests a stream or an orbiting 
GC system, but clearly this could also be a projection effect. While difficult to obtain, large
kinematic sample of GCs could be used to compare the observed spatial distribution of GCs with 
the results from simulations of the formation of elliptical galaxies, providing a way to constrain the 
dynamical evolution of the host galaxy (see, for example~\citealt{bekki2005}).

\section{Conclusions}
\label{sec:conclusions}

We have developed a new approach to the study of the 2D distribution of GCs in elliptical galaxies. The 
method is an implementation of the KNN method presented in~\cite{dressler1980}, supplemented by 
Monte-Carlo simulations to establish the statistical significance of the results. We have applied this 
method to NGC4261, a ``test galaxy'' where significant 2D anisotropy in the GC distribution has been 
reported (B+12). We confirm that the 2D distribution of GC is not azimuthally isotropic. 
Moreover, we demonstrate that the 2D distribution departures from the average GC radial distribution 
results in highly significant spiral-like or broken shell features. While the southern feature is seen in 
both red and blue GC subsamples, the northern anisotropy shows some evidences of segregation 
of red and blue GCs. The good characterization of the 2D distribution of GC that we have achieved 
provides a new benchmark for future simulations of galaxy merging evolution.

\acknowledgements

R. D'Abrusco gratefully acknowledges the financial 
support of the US Virtual Astronomical Observatory, which is sponsored by the
National Science Foundation and the National Aeronautics and Space Administration.
TF acknowledges support from the CfA and the ITC prize fellowship programs. This work 
was partially supported by the {\it Chandra} X-ray Center (CXC), which is operated by the 
Smithsonian Astrophysical Observatory (SAO) under NASA contract NAS8-03060.

{}

\end{document}